\documentclass[notitlepage, final, 10pt, prl, aps, twocolumn, floatfix, letter, longbibliography]{revtex4-2}

\pdfoutput=1

\usepackage[utf8]{inputenc}
\usepackage[english]{babel}
\usepackage{lmodern}
\usepackage{xcolor}
\definecolor{navyblue}{rgb}{0.0, 0.0, 0.5}
\usepackage{amsmath, amssymb, amsfonts, bm}
\usepackage{latexsym}
\usepackage{bbm}
\usepackage{mathtools}
\usepackage{placeins}
\usepackage{braket}
\usepackage{booktabs}
\usepackage[protrusion=true, expansion=true]{microtype}
\usepackage[colorlinks=true, breaklinks=false, linkcolor=navyblue, urlcolor=navyblue, citecolor=navyblue]{hyperref}

\usepackage{graphicx}
\usepackage{float}
\usepackage{siunitx}

\DeclareMathOperator{\Tr}{Tr}

\begin{document}
\title{The phases of the disordered Bose--Hubbard model with attractive interactions}
\author{Olli Mansikkamäki}
\author{Sami Laine}
\author{Matti Silveri}
\affiliation{Nano and Molecular Systems Research Unit, University of Oulu, P.O.\ Box 3000, FI-90014 University of Oulu, Finland}
\date{\today}
\begin{abstract}We study the quantum ground state phases of the one-dimensional disordered Bose--Hubbard model with attractive interactions, realized by a chain of superconducting transmon qubits or cold atoms. We map the phase diagram using perturbation theory and exact diagonalization. Compared to the repulsive Bose--Hubbard model, the quantum ground state behavior is dramatically different. At strong disorder of the on-site energies, all the bosons localize into the vicinity of a single site, contrary to the Bose glass behavior of the repulsive model. At weak disorder, depending on hopping, the ground state is either superfluid or a W state, which is a multi-site and multi-particle entangled superposition of states where all the bosons occupy a single site. We show that the robustness of the W phase against disorder diminishes as the total number of bosons increases. 
\end{abstract}
\maketitle

\paragraph*{Introduction} The Bose--Hubbard model is a paradigmatic model of quantum matter and quantum phase transitions, with applications ranging from magnetism to disordered superfluid helium~\cite{girvin19, greiner_quantum_2002, stoferle_transition_2004, bloch_many-body_2008}. It is canonically characterized by a repulsive boson-boson interaction disfavoring local multi-occupancy, together with boson hopping which models excitation kinetics. When repulsive interaction dominates hopping, the ground state is the Mott insulating phase where a fixed integer number of bosons are located on each lattice site. Otherwise, the ground state is the delocalized superfluid~\cite{kuhner_one-dimensional_2000, cazalilla_one_2011}. In the presence of disorder, a third phase\----the Bose glass\----emerges between the Mott insulator and the superfluid phases~\cite{giamarchi_anderson_1988, fisher_boson_1989, haviland_onset_1989,hebard_magnetic-field-tuned_1990, rapsch_density_1999, fallani_ultracold_2007, cazalilla_one_2011,yu_bose_2012, meldgin_probing_2016}. The Bose glass is an insulating phase with finite compressibility caused by disorder localization. 

The attractive Bose--Hubbard model has remained much less studied than its repulsive counterpart. At strong disorder, many-body localization emerges~\cite{roushan_spectroscopic_2017, orell_probing_2019, chiaro_direct_2020} for highly-excited states. The quantum ground state behavior changes dramatically when switching from repulsive to attractive interaction~\cite{dorignac_quantum_2004, jack_bose-hubbard_2005, buonsante_attractive_2005, oelkers_ground-state_2007, buonsante_quantum_2010}. 
When attractive interaction dominates hopping, in the absence of disorder, the ground state is the~W~state, a fascinating self-trapped multiparticle entangled state comprising a cat-state-like superposition of states with all the bosons occupying a single site~\cite{bernstein_1990, sorensen_relative_2012, gangat_deterministic_2013}. However, the interplay of disorder and attractive interactions has not been studied before for the quantum ground states.

The attractive Bose--Hubbard model is an important model for arrays of superconducting transmon devices, a leading platform for large-scale quantum science experiments. A transmon is an anharmonic bosonic oscillator with negative anharmonicity~\cite{koch_charge-insensitive_2007}. Fabrication disorder~\cite{underwood_low-disorder_2012} has hindered their utilization in large-scale quantum simulators~\cite{houck_-chip_2012,dalmonte_realizing_2015} of other than disorder physics~\cite{roushan_spectroscopic_2017,xu_emulating_2018, zha_ergodic-localized_2020, chiaro_direct_2020,gong_experimental_2020}. The size of experimentally demonstrated transmon arrays has grown rapidly from a few to over \num{50} sites~\cite{hacohen-gourgy_cooling_2015, ma_dissipatively_2019, xu_emulating_2018, reagor_demonstration_2018, arute_quantum_2019,zha_ergodic-localized_2020,campbell_universal_2020,  guo_stark_2020, gong_experimental_2020}. Thus, an array of coupled transmons realizes the disordered attractive Bose--Hubbard model in a natural manner~\cite{hacohen-gourgy_cooling_2015, orell_probing_2019}. Furthermore, the attractive Bose--Hubbard model is immediately applicable also for cold atoms in optical lattices, where the interaction can be tuned from repulsive to attractive via the Feshbach resonance~\cite{theis_tuning_2004, bloch_many-body_2008}. 

In this letter, we use exact diagonalization and perturbation theory to construct the ground state phase diagram of the one-dimensional disordered attractive Bose--Hubbard model, and provide analytical expressions for the states belonging to the W phase, the superfluid phase, and the localized phase. Our main result is that the robustness of the W phase against disorder diminishes exponentially as the total number of bosons is increased. Finally, we propose a possible realization of these phases using transmon chains with experimentally feasible parameters.

We note that phases and phase transitions are, strictly speaking, only defined in the thermodynamic limit, that is, when both the number of lattice sites and the number of particles approach infinity. For the Bose--Hubbard model with attractive interactions, this is ill-defined since the two limits are non-commutative~\cite{oelkers_ground-state_2007} and the ground state energy is not bounded below. However, seeing that the finite-size behavior of the model resembles that of a system with well defined phases, the concept of phase is frequently used~\cite{oelkers_ground-state_2007, deng_quantum_2008, gangat_deterministic_2013}.

\begin{figure*} 
\includegraphics[width=0.85\linewidth]{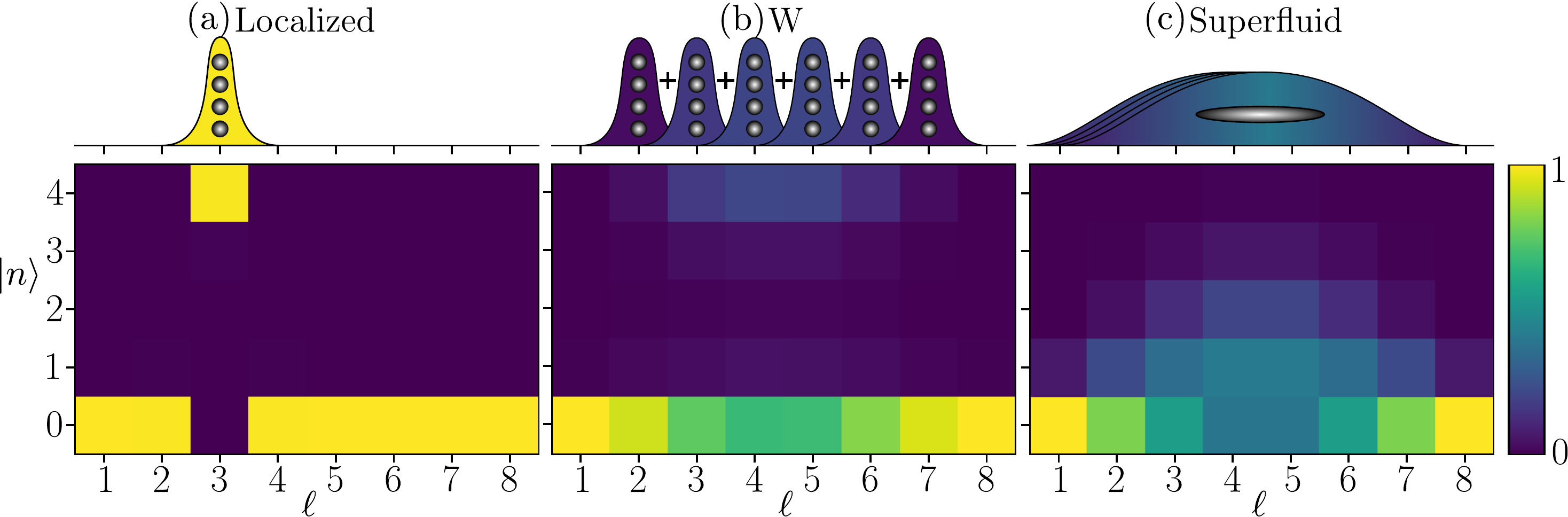}
\caption{\label{fig:prob} A schematic illustration and the local occupation density $p_{n\ell}=\braket{n|\hat \rho_\ell|n}$ as a function of the transmon site $\ell$ and the local Fock state $\ket{n}$ for (a)~the~localized state, (b)~the~W~state, and (c)~the~superfluid~state at the scaled hopping frequency $\tau = 0.05$, $0.15$, and $1$, respectively. A single disorder realization at the scaled disorder strength $\delta =0.33\cdot10^{-3}$ was used. The local density matrix $\hat \rho_\ell$ is calculated by tracing over all the other sites: $\hat \rho_\ell=\Tr_{\{i\neq \ell\}}\left(\ket{\psi}\bra{\psi}\right)$, where $\ket{\psi}$ is the ground state. }
\end{figure*}

\paragraph*{Model} The disordered attractive Bose--Hubbard model with $L$ sites is defined in the basis of the local bosonic annihilation $\hat a_\ell$, creation $\hat a_\ell^\dag$, and occupation number $\hat{n}_\ell =  \hat{a}^\dagger_\ell \hat{a}^{}_\ell$ operators by the Hamiltonian~\cite{hacohen-gourgy_cooling_2015, roushan_spectroscopic_2017, ma_dissipatively_2019, orell_probing_2019}
\begin{equation}
    \frac{\hat{H}}{\hbar} = \sum_{\ell = 1}^{L}\left[ \omega_\ell\hat{n}_\ell - \frac{U}{2} \hat{n}_\ell (\hat{n}_\ell - 1)+ J\left(\hat{a}^\dagger_{\ell+1} \hat{a}^{\phantom{\dagger}}_{\ell} +\hat{a}^{\dagger}_{\ell} \hat{a}^{\phantom{\dagger}}_{\ell+1} \right)\right]. \label{eq1}
\end{equation}
Here, $\omega_\ell$ represent random disorder of on-site energies. We draw them from a uniform distribution in the interval $\omega_\ell\in [-D, D]$, with $D$ being the strength of the disorder. Since the model conserves the total number of bosons $\hat N=\sum_{\ell=1}^L\hat n_\ell$, we consider a fixed $N$ and can thus ignore the mean on-site energy. The strength of the attractive interactions is given by $U$, and $J$ is the hopping frequency. Finally, $\hbar$ is the reduced Planck's constant. The bounded spectrum at fixed $N$ implies that the most excited eigenstate of the Hamiltonian~\eqref{eq1} is the quantum ground state of the corresponding repulsive model, and vice versa. For the sake of experimental relevance, we focus on open chains with $\hat a_{L+1}=0$ in Eq.~\eqref{eq1} (see Ref.~\onlinecite{suppl} for the corresponding results for a periodic chain). In what follows, we measure all energies in units of the characteristic energy $\hbar U N (N-1)$, yielding the scaled energy $\varepsilon=E/\hbar U N (N-1)$, hopping frequency $\tau=J/U (N-1)$, disorder strength $\delta=D / U(N-1)$, and on-site energies $\sigma_\ell = \omega_\ell / U(N-1)$.

\paragraph*{Localized phase} Let us first consider strong attractive interactions and strong disorder $\delta \gg 1/(N-1) \gg \tau$. The attractive interactions force all the bosons to occupy the site $\ell_0$ with the lowest on-site energy $\sigma_{\ell_0}$, leading to total energy $\varepsilon_{\rm loc}^0 = - 1/2 + \sigma_{\ell_0}$. With strong disorder and ignoring hopping, the ground state is a  product state of the form
\begin{equation}
  \ket{\psi^{0}_{\ell_0}} = \ket{n_{\ell_0} = N}, \label{eq.loc0}   
\end{equation}
where $\ket{n_{\ell_0}=N}$ denotes the state where $N$ bosons occupy the site $\ell_0$ and other sites are empty. We refer to this as the localized state, not to be confused with Anderson localization of the non-interacting situation or the many-body localization of the highly-excited states~\cite{roushan_spectroscopic_2017, orell_probing_2019, chiaro_direct_2020}. When we take into account the hopping up to first order in non-degenerate perturbation theory, the localized state $\ket{\psi^{0}_{\ell_0}}$ gets a correction of the form
\begin{equation}
    \ket{\widetilde{\psi}^{1}_{\ell_0 \pm 1}} = \tau \sqrt{N} \frac{\ket{n_{\ell_0 \pm 1}=1, n_{\ell_0}=N-1}}{(\sigma_{\ell_0}-\sigma_{\ell_0 \pm 1}) - 1},
\end{equation}
that is, a state localized onto the site $\ell_0$ will be
\begin{equation}
\ket{\psi^{1}_{\ell_0}} =\ket{\psi^{0}_{\ell_0}} + \ket{\widetilde{\psi}^{1}_{\ell_0 + 1}} + \ket{\widetilde{\psi}^{1}_{\ell_0 - 1}},\label{eq.loc1}
\end{equation}
up to a normalization factor [see Fig.~\ref{fig:prob}(a)]. The same form applies to every localized state excepting the ends, where only a single correction is added. After averaging over the disorder, the second-order energy of the localized ground state phase is
\begin{align}
  \varepsilon_{\rm loc} &=-\frac{1}{2}- \delta \frac{L-1}{L+1}(1-2\tau^2)-2\tau^2\frac{L-1}{L}. \label{eq.locener}
\end{align}
When reducing disorder, the ground state phase changes either into the superfluid phase or into the W phase, depending on hopping. 

\paragraph*{W phase}Let us next consider weak hopping, $\tau\ll 1$, in the absence of disorder. Then the states $\ket{n_{\ell=1,\ldots,L}=N}$ with the energy $\varepsilon^0=-1/2$ are degenerate and coupled by a weak high-order hopping interaction, yielding a quantum ground state which is a superposition of the localized states $\ket{\psi_\ell^1}$ of Eq.~\eqref{eq.loc1}, see Fig.~\ref{fig:prob}(b). This state is called the W state \cite{heaney_extreme_2011, wang_nonlocality_2013,gangat_deterministic_2013} because of its resemblance with the W state of entangled qubits~\cite{dur_three_2000}. 

To solve the W state analytically, we resort to high-order degenerate perturbation theory~\cite{sakurai2017}, detailed in Ref.~\onlinecite{suppl}. First, the sites at the ends of the open chain have only one neighbor each. Thus, the effective hopping energy near the ends is higher than elsewhere, reducing the number of degenerate states by $\ell_{\rm s}=\lceil N/2 \rceil - 1$ counting from both ends of the chain. For simplicity, let us consider even $L$ and odd $N$, such that the number of degenerate states $L_{\rm d}=L-2\ell_{\rm s} \ge 2$. Then the degenerate perturbation theory in the $N$th order shows that the remaining states $\ket{n_\ell=N}$ are each coupled with their neighboring-site counterparts, which gives us the superposition [see Fig.~\ref{fig:prob}(b)]

\begin{equation}
  \ket{\psi_{\rm w}^1}=\sqrt{\frac{2}{L_d+1}}\sum_{\ell=1}^{L_{\rm d}} (-1)^\ell \sin \left( \frac{\pi \ell }{L_d + 1} \right) \ket{\psi^1_{\ell+\ell_{\rm s}}}.  \label{eq.W0} 
\end{equation}

The existence of the W state requires a non-zero hopping frequency, which is why we use the first-order perturbed states of Eq.~\eqref{eq.loc1} in the superposition. The shape of the W state depends strongly on the total density $N / L$ of the bosons--- the higher the density, the more the bosonsbunch toward the middle of the chain, forming a self-trapping state~\cite{buonsante_quantum_2010}. For even $L$, the limit is an equal superposition of the two middlemost localized states of Eq.~\eqref{eq.loc1}. For odd $L$, the limit is just the middlemost localized state. The mean second-order energy of the W state is
\begin{equation}
    \varepsilon_{\rm W}=-\frac 1 2 -2\tau^2,\label{eq.wener}
\end{equation}
with no disorder contribution after averaging over different realizations.

When introducing disorder, we can probe the weak interaction between the degenerate states formed through high-order bosonic hopping. The magnitude of the $N$th order hopping interaction between the state $\ket{n_\ell=N}$ and the states $\ket{n_{\ell\pm 1}=N}$ is $[\alpha(N) \tau]^N$, where $\alpha(N) = \left[(N - 1)^{N - 1}/(N - 1)! \right]^{1/N}$ is a coefficient dependent on the boson number. At large $N$, the value of $\alpha(N)$ approaches Euler's number $e$. The magnitude of disorder needs to be of the same order or stronger than this effective hopping energy to cause the W state to disintegrate into a localized state. To sum up, at weak disorder and hopping, the W phase can exist only when $\delta \lesssim [\alpha(N) \tau]^N$, showing that the robustness of the W phase against disorder diminishes exponentially with increasing total boson number.

\paragraph*{Superfluid phase} When hopping dominates interactions, $\tau \gg 1$, the bosons can move from site to site largely unhindered by each other, that is, the ground state is a superfluid. It is easier here to work in the reciprocal space, i.e., in the eigenbasis of the hopping term of the Hamiltonian, accessible via the transformation
\begin{equation}
    \hat{c}_k = \sqrt{\frac{2}{L + 1}} \sum_{\ell = 1}^L \sin \left( \frac{\pi \ell k}{L + 1} \right) \hat{a}_\ell \label{eq.recip}
  \end{equation}
in the open chain.
Analogously to the spatially localized states of Eq.~\eqref{eq.loc0}, we may form localized states in the reciprocal space,
\begin{equation}
  \ket{\psi^{0}_k} = \ket{\eta^{}_{k} = N}, \label{eq.sf0}
\end{equation}
where $\ket{\eta^{}_{k}=N}$ denotes the state where $N$ bosons occupy the $k$th reciprocal mode and other modes are empty. In the limit of vanishing interactions and disorder, the superfluid ground state $\ket{\psi^{0}_{\rm SF}}=\ket{\psi_{k=L}^{0}}$ is completely localized to the lowest-energy mode of the reciprocal space, $k = L$, with the energy $\varepsilon_{\rm SF}^{0}=-2 \tau \cos[\pi/(L+1)]$. Although the excitations created via $\hat{c}^\dagger_k$ are localized in the reciprocal space, they are delocalized in the position space, see Fig.~\ref{fig:prob}(c) and Ref.~\onlinecite{suppl} for visualizations. Both the interaction term and the disorder term have a weakly delocalizing effect in the reciprocal space, similar to the effect of the hopping term on the spatially localized state. The ground state can be calculated by first transforming these into the reciprocal basis and considering their effects perturbatively~\cite{suppl}. The second-order disorder-averaged energy is given by
\begin{equation}
  \varepsilon_{\rm SF}=-2 \tau \cos \left ( \frac{\pi}{L+1} \right)-\frac{3}{4 (L+1)}- \frac{a}{\tau}\left(\delta^2+b\right), \label{eq.sfener}
\end{equation}
where $a$ and $b$ are coefficients dependent on $N$ and $L$ (at $N=4$ and $L=8$, $a\approx 0.23$ and $b \approx 0.05$, see Ref.~\onlinecite{suppl}). Notice that this condensate of bosons can indeed be called a superfluid since it is able to support metastable persistent currents in a periodic chain \cite{arwas2015, arwas2017}.

\paragraph*{Phase diagram} Analytical results indicate that we could expect ground state phase transitions at parameters where the energies of any two phases [Eqs.~\eqref{eq.locener},~\eqref{eq.wener}, and~\eqref{eq.sfener}] are close. To study this in detail, we calculate the quantum ground state of the Hamiltonian~\eqref{eq1} by numerical exact diagonalization. To identify all three phases in numerical calculations, we need at least two indicator quantities. We can use the fact that the states are characterized by different types of localization: the localized and superfluid phases are localized in spatial and reciprocal bases, respectively, and the W state is a superposition of spatially localized states. One quantity measuring the degree of localization is the inverse participation ratio
\begin{equation}
  \mathcal{P}_{\rm s/r} = \frac{1}{L - 1}\left(\frac{N^2}{\sum_{m = 1}^L|\braket{\psi|\hat n^{\rm s/r}_m|\psi}|^2} - 1\right),\label{eq.P}
\end{equation}
which we can calculate in both the position space $\mathcal{P}_{\rm s}$ and in the reciprocal space $\mathcal{P}_{\rm r}$ separately. Here, $\hat n^{\rm s}_m = \hat a_m^\dag\hat a^{}_m$ and $\hat n^{\rm r}_m =\hat c_m^\dag\hat c^{}_m$ are the number operators in the corresponding spaces. The inverse participation ratio yields zero if the state is localized and one if the state is completely delocalized. We can use $\mathcal{P}_{\rm s}$ to distinguish the spatially localized states from the superfluid states and the W~states. Conversely, we can use $\mathcal{P}_{\rm r}$ to distinguish the superfluid states from the W~states and the localized states.

\begin{figure}
    \centering
    \includegraphics[width=\columnwidth]{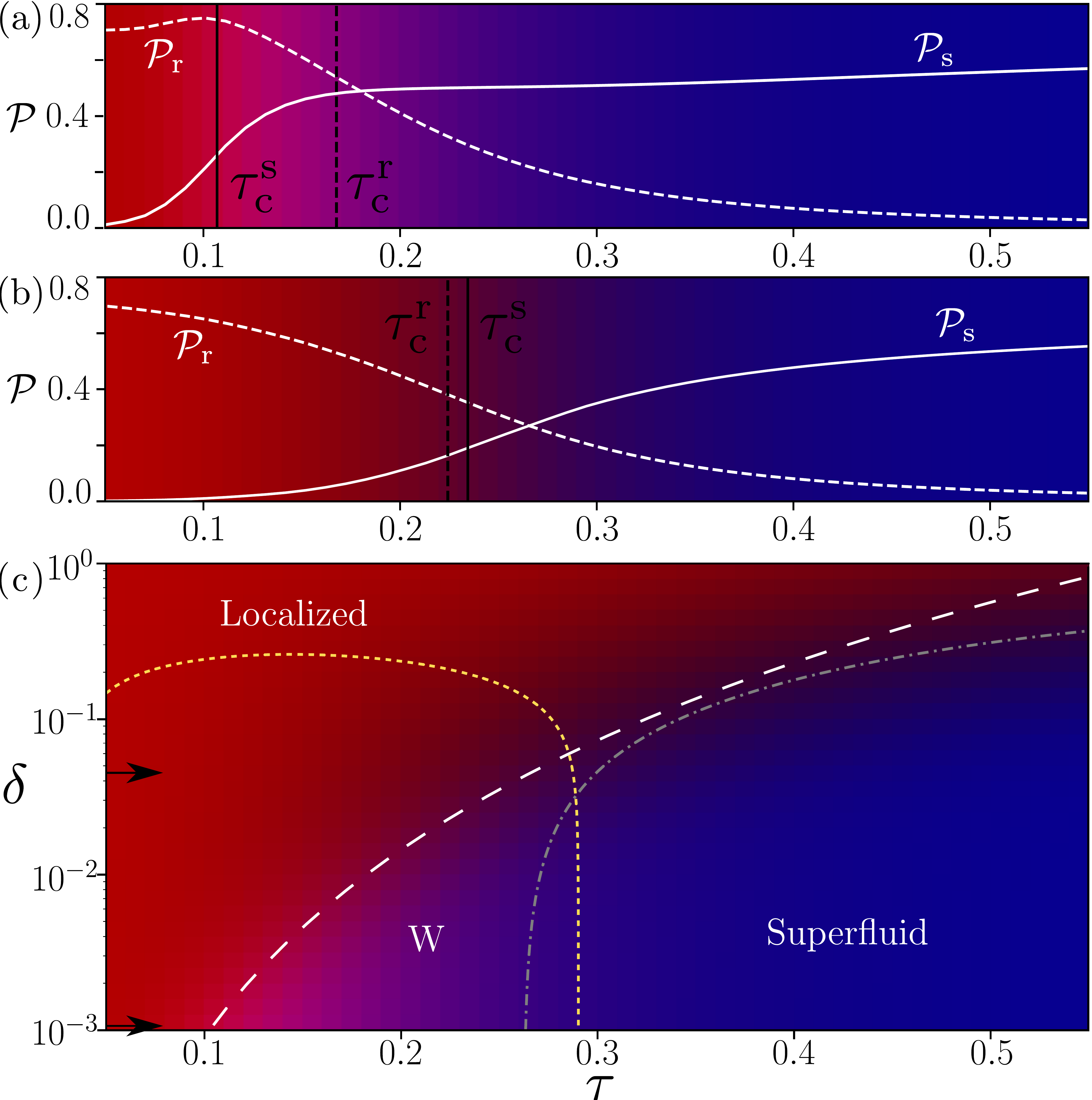}
    \caption{(a) and (b) In white, the inverse participation ratios $\mathcal{P}_{\rm s}$ (solid) and $\mathcal{P}_{\rm r}$ (dashed), defined in Eq.~(\ref{eq.P}), of the quantum ground state of the disordered attractive Bose--Hubbard model (\ref{eq1}) as a function of the scaled hopping frequency $\tau$ at two values of the scaled disorder strength, (a) $\delta = 0.001$ and (b) $\delta = 0.036$. In black, the corresponding critical scaled hopping frequencies $\tau^{\rm s/r}_{\rm c}$ defined in Eq.~(\ref{eq.tau_c}).
    (c) The two inverse participation ratios $\mathcal{P}_{\rm s/r}$, represented as a single colormap, as a function of the scaled hopping frequency $\tau$ and the disorder strength $\delta$. The amount of blue (red) indicates the value of $\mathcal{P}_{\rm s}$ ($\mathcal{P}_{\rm r}$): The bluer (redder) the pixel, the more delocalized the ground state is spatially (in the reciprocal space). When both of the inverse participation ratios are non-zero, the color of the corresponding pixel is purple, indicating the W phase. The overlaid lines indicate the phase boundaries estimated from the analytical results: localized-to-W at $\delta=2( \alpha \tau)^N$ (white dashed line), superfluid-to-localized at $\varepsilon_{\rm SF}=\varepsilon_{\rm loc}$ (gray dash-dotted line), and  W-to-superfluid at $\varepsilon_{\rm W}=\varepsilon_{\rm SF}$ (yellow dotted line). The arrows indicate the locations of the horizontal cut-offs shown in (a) and (b). The ground states were numerically computed for an open chain of length $L=8$ with the total number of bosons $N=4$, and averaged over $1000$ disorder realizations. See Ref.~\cite{suppl} for the results for corresponding periodic chains. Note that $\tau \geq$ 0.05 in this and all the following figures.}
    
    \label{fig2}
\end{figure}

Keeping in mind the theoretical limitations of increasing the total boson number $N$ mentioned in the introduction, we are nevertheless interested in the  phase boundaries of the three phases as the system size in $N$ is increased, which physically means approaching the semiclassical limit. For this purpose, we define the critical scaled hopping frequency~\cite{buonsante_ground-state_2007}
\begin{equation}
    \tau^{\rm s/r}_{\rm c}=\mathop{\rm arg max}_{\tau}\left|\frac{\partial\, \mathcal{P}_{\rm s/r}}{\partial\, \tau} \right| \label{eq.tau_c}
\end{equation}
to locate the transition point from localized to delocalized state both in the position space ($\tau^{\rm s}_{\rm c}$) and in the reciprocal space ($\tau^{\rm r}_{\rm c}$).

In Fig.~\ref{fig2}(a), we show the inverse participation ratios $\mathcal P_{\rm s/r}$ for the ground states of the Hamiltonian~\eqref{eq1} as a function of the scaled hopping frequency $\tau$ at weak disorder. The critical scaled hopping frequencies $\tau^{\rm s/r}_{\rm c}$ indicate that the system localizes in the position space (solid) when $\tau < \tau^{\rm s}_{\rm c} \approx 0.11$ and in the reciprocal space (dashed) when $\tau > \tau^{\rm r}_{\rm c} \approx 0.17$. In the region between, we identify the W state which is delocalized spatially but not yet localized in the reciprocal space. Note that~$\mathcal{P}_{\rm r}$~attains its maximum at a non-zero $\tau$, within the W phase. This stems from the fact that, due to the shape of the transformation~\eqref{eq.recip}, a superposition of spatially localized states can be more uniformly distributed in the reciprocal space than a single localized state. However, as the number of bosons $N$ grows, the W state gets more and more localized towards the middle of the chain, and thus the local maximum of $\mathcal{P}_{\rm r}$ at $\tau > 0$ disappears. At moderate disorder, depicted in Fig.~\ref{fig2}(b), increasing hopping simultaneously delocalizes the system in the position space and localizes it in the reciprocal space, with a crossover at $\tau\approx 0.23$, indicating absence of the~W~state.

\begin{figure}
    \centering
    \includegraphics[width=\columnwidth]{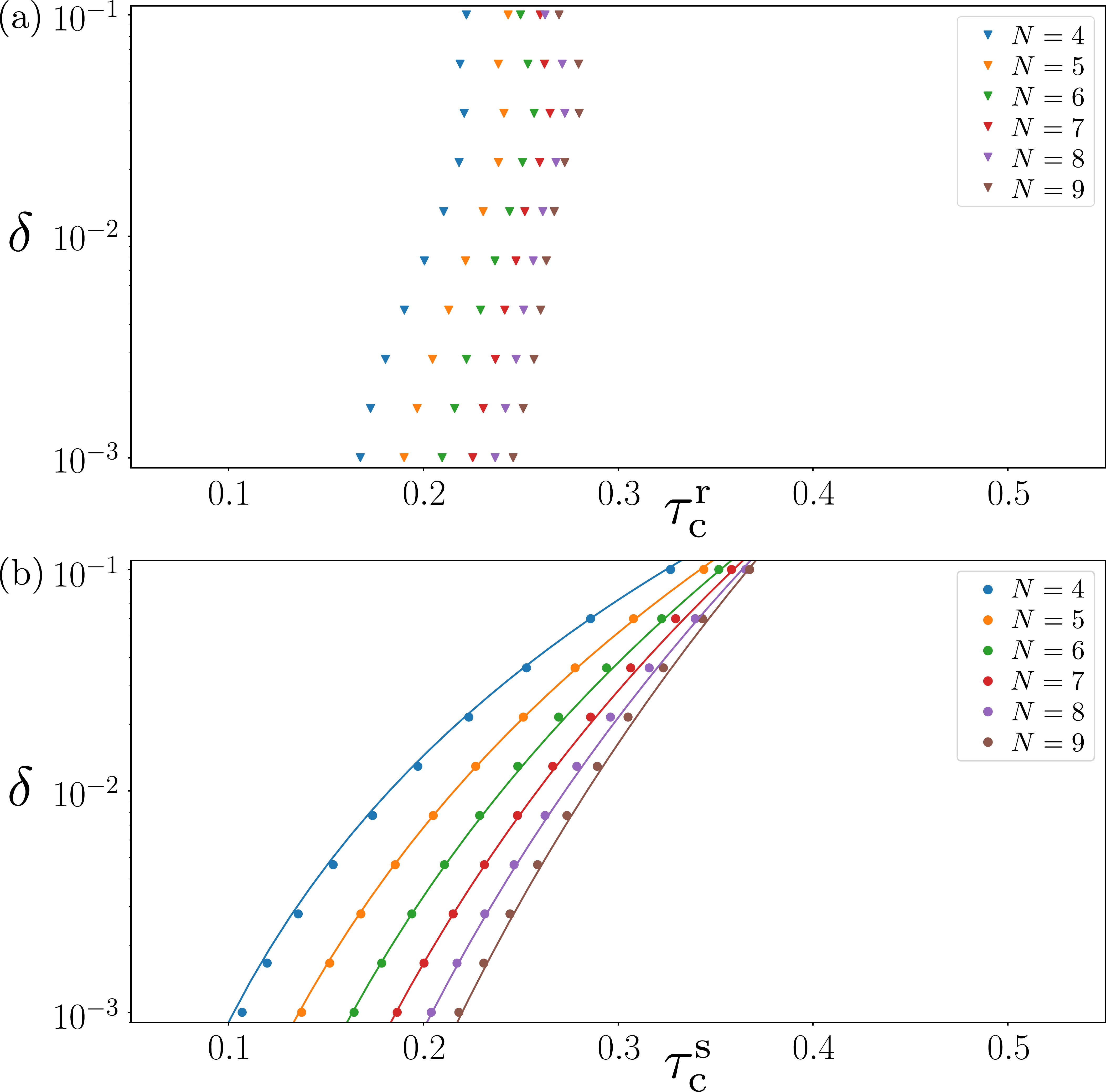}
    \caption{The critical scaled hopping frequencies (a) $\tau^{\rm r}_{\rm c}$ (colored filled triangles) and (b) $\tau^{\rm s}_{\rm c}$ (colored filled circles), defined in Eq.~(\ref{eq.tau_c}), as a function of the total boson number $N=4, \ldots, 9$ and the scaled disorder strength $\delta$. The chain length is $L=8$ and the results are averaged over \num{2000} disorder realizations. The analytically predicted phase boundary between the localized phase and the W phase is shown as colored lines at $\delta=2[\alpha(N)\tau_{\rm c}^{\rm s}]^N$ in (b).}
    \label{fig3}
\end{figure}

To explore the ground state phases in the full parameter space, we have computed the inverse participation ratios $\mathcal{P}_{\rm s/r}$ as a function of both the scaled hopping frequency $\tau$ and the scaled disorder strength $\delta$. These are represented in Fig.~\ref{fig2}(c) as a colormap, where the amount of blue (red) color in each pixel indicates the value of $\mathcal{P}_{\rm s}$ ($\mathcal{P}_{\rm r}$). In the W phase, both indicators are non-zero, shown with purple color. Notice that with $N=4$ and $L=8$, the W phase exists only at $\delta \lesssim 0.03$. At stronger disorder, there is a direct transition from the localized phase into the superfluid phase. The overlaid lines of Fig.~\ref{fig2}(c) indicate the phase boundaries predicted by equating the analytical ground state energies. 

At strong disorder, the localized phase changes into the superfluid phase near the curve $\varepsilon_{\rm loc}=\varepsilon_{\rm SF}$ (dash-dotted line). At weak disorder, the superfluid phase changes into the W phase in the proximity of the curve $\varepsilon_{\rm SF}=\varepsilon_{\rm W}$ (dotted line), while the W phase disintegrates into the localized phase at $\delta=2[\alpha(N)\tau]^N$ (dashed line). Here, the numerical coefficient $2$ in the localized-to-W boundary was obtained by fitting the analytically predicted expression (see Ref.~\onlinecite{suppl}) $\delta=(3A/2)[\alpha(N)\tau]^N$, with $A$ a constant of the order of unity, to the numerical data.

Figures~\ref{fig3}(a--b) show the critical hopping frequencies $\tau^{\rm r}_{\rm c}$ and $\tau^{\rm s}_{\rm c}$, respectively, as a function of the disorder strength $\delta$ for varying total boson number $N$. In Fig.~\ref{fig3}(a), the transition between the W phase and the superfluid phase shifts towards larger hopping frequency as $N$ is increased. We can explain this qualitatively through the fact that the coupling between the reciprocal modes by the interaction term of the Hamiltonian~\eqref{eq1} increases with $N$, contained in the coefficients $a$ and $b$ of Eq.~\eqref{eq.sfener}. In Fig.~\ref{fig3}(b), the data points indicate the transition between the localized and the W phase. Here, we observe clearly that the W phase exists only at  $\delta \lesssim 2[\alpha(N)\tau]^N$ (solid colored lines), confirming our analytical prediction.

\paragraph{Realization with transmon arrays} With transmon arrays, the hopping term is realized by the capacitive coupling between the transmons~\cite{koch_charge-insensitive_2007, dalmonte_realizing_2015}, resulting in the value of $J / 2 \pi $ ranging between \SI{10}{\mega\hertz} and \SI{100}{\mega\hertz}. The interaction term originates from an approximation of the cosinusoidal potential of the Josephson junctions of the transmons~\cite{koch_charge-insensitive_2007}, with typical values of $U / 2\pi $ between \SI{200}{\mega\hertz} and \SI{300}{\mega\hertz}. The disorder $\omega_\ell$ models the small unintentional variations in manufactured devices, but it can also be artificially magnified or reduced by applying magnetic flux through the loop formed by the two parallel Josephson junctions of the transmons~\cite{koch_charge-insensitive_2007, hacohen-gourgy_cooling_2015, roushan_spectroscopic_2017, ma_dissipatively_2019, orell_probing_2019, arute_quantum_2019}, yielding $D/2\pi$ in the range from \SI{100}{\kilo\hertz} to \SI{2}{\giga\hertz}. The experimentally realistic range of parameters belongs roughly to the interval $J / U \in [0.03, 0.5]$ and $ D/U \in [10^{-4}, 10^2]$, allowing the realization of all the ground state phases of the disordered attractive Bose--Hubbard model. For modern transmons, the rate $\Gamma_1$ of losing bosons (photons) is low~\cite{ma_dissipatively_2019, gong_experimental_2020, campbell_universal_2020}, with $\Gamma_1/2\pi$ in the range of a few~kHz. This should be contrasted with the effective hopping frequencies in the system. For example, in the W phase, the ratio between the effective hopping frequency $\widetilde{J} = U N (N-1) [\alpha(N) \tau]^N$  and the many-body dissipation rate is $\widetilde{J}/(N\Gamma_1) \gtrsim 50$ with the parameters of Ref.~\onlinecite{campbell_universal_2020}, indicating an ample window of coherent dynamics to form and detect the W phase before a disruptive photon loss event.

The Hamiltonian~\eqref{eq1} becomes increasingly worse a model of a transmon array as the number of bosons on a single site increases, since the interaction term $-U\hat n(\hat n-1)/2$ is just the lowest anharmonicity term of the cosine potential of the Josephson junction~\cite{koch_charge-insensitive_2007}.
However, a more fundamental limitation is that the cosine potential also implies that a transmon has only a finite number of discrete bound states~\cite{pietikainen_photon_2019}. This number for modern transmon parameters is $\omega_0/(\sqrt{8}U)\approx 10$, with $\omega_0$ the mean on-site energy. Adding more bosons will change the transmon spectrum and couplings, rendering the Bose--Hubbard model insufficient. This warrants our choice of $N<10$.

To detect and distinguish the state, the population of every transmon or a subset of them can be measured with high accuracy by coupling them individually to dispersive readout resonators~\cite{Krantz19}, as demonstrated recently in many-body settings~\cite{ma_dissipatively_2019, roushan_spectroscopic_2017}. This way, one can measure the local occupation density that uniquely identifies each phase as shown in Fig.~\ref{fig:prob}. By utilizing the sophisticated driving protocols, one can realize cooling and stabilization schemes that can be used to achieve the quantum ground states at fixed boson numbers, as was experimentally demonstrated in the seminal works~\cite{hacohen-gourgy_cooling_2015, ma_dissipatively_2019}.

\paragraph*{Conclusions} In this work, we have concentrated on the static ground state properties of the attractive Bose--Hubbard model with weak to moderate energy disorder. We mapped the ground state phase diagram and focused on the W phase, which can be experimentally realized with transmon arrays only within stringent bounds for the disorder and coupling strengths. Seeing that decoherence and dissipation are common issues in superconducting quantum devices, and that the loss of bosons can change the nature of the ground state, even a system starting from the ground state for a given number of bosons $N$ may experience quite interesting dynamics. Specifically, even if the bare system parameters $J$ and $U$ are not changed, boson losses due to dissipation effectively act as quenches in the scaled hopping frequency $\tau=J/U(N-1)$ and disorder strength $\delta=D/U(N-1)$. This allows for the exploration of the parameter space $(\tau, \delta)$, although only in discrete steps. A more detailed discussion of the dynamics due to dissipation and quenches in the parameters of the model, including a potential realization of dynamical quantum phase transition~\cite{heyl_dynamical_2018, lacki_dynamical_2019}, is an interesting subject of future work. Another intriguing question is the structure of the ground state phase diagram in two and higher dimensions, relevant even for current transmon arrays~\cite{arute_quantum_2019}.

\paragraph*{Acknowledgments}
  The authors would like to thank Steven Girvin and Tuure Orell for useful discussions on this and other closely related topics. This research work was financially supported by the Alfred Kordelin foundation, the Emil Aaltonen foundation, the Kvantum Institute of the University of Oulu, and the Academy of Finland under Grants No. 316619 and No. 320086. The numerics are implemented in Julia~\cite{bezanson2017julia} using the KrylovKit and Optim~\cite{mogensen2018optim} packages.

\bibliography{refs.bib}{}

\end{document}


\title{Supplementary material: The phases of the disordered Bose--Hubbard model with attractive interactions}
\author{Olli Mansikkamäki}
\author{Sami Laine}
\author{Matti Silveri}
\affiliation{Nano and Molecular Systems Research Unit, University of Oulu, P.O.\ Box 3000, FI-90014 University of Oulu, Finland}
\date{\today}

\maketitle

\section{Model}
In this supplementary material, we take a closer look at the asymptotic phases of the attractive Bose--Hubbard model with $L$ sites. We consider both an open chain and a periodic chain, the former of which was discussed in the main text. In both cases, the Hamiltonian of the system can be written as a sum of four parts,
\begin{equation}
\hat{H} = \hat{H}_M + \hat{H}_D + \hat{H}_U + \hat{H}_J,
\end{equation}
where
\begin{align}
\hat{H}_M / \hbar &= \omega_0 \sum_{\ell = 1}^{L} \hat{n}_\ell, \\
\hat{H}_D / \hbar &= \sum_{\ell = 1}^{L} \omega_\ell \hat{n}_\ell, \\
\hat{H}_U / \hbar &= - \frac{U}{2} \sum_{\ell = 1}^{L} \hat{n}_\ell (\hat{n}_\ell - 1), \\
\hat{H}_J / \hbar &= J \sum_{\ell = 1}^{L}\left(\hat{a}^\dagger_{\ell+1} \hat{a}^{}_{\ell} + \hat{a}^{\phantom{\dag}}_{\ell+1} \hat{a}^{\dagger}_{\ell} \right).
\end{align}
Here, $\hat a_\ell^\dag$ and $\hat a_\ell$ are the bosonic creation and annihilation operators at site $\ell$ satisfying the commutation relations $[\hat{a}_\ell^\dagger, \hat{a}_m^\dagger] = [\hat{a}_\ell, \hat{a}_m] = 0$, $[\hat{a}_\ell, \hat{a}_m^\dagger] = \delta_{\ell, m}$, with $\delta_{\ell, m}$ the Kronecker delta, and $\hat{n}_\ell =  \hat{a}^\dagger_\ell \hat{a}^{}_\ell$ is the corresponding occupation number operator. For an open chain we define $\hat a_{L+1}^\dag = \hat a_{L+1} = 0$, while for a closed chain we set $\hat a_{L+1}^\dag = \hat a_1^\dag$ and $\hat a_{L+1} = \hat a_1$.

The first two terms in the Hamiltonian, $\hat{H}_M$ and $\hat{H}_D$, represent the on-site energies of the bosons neglecting interactions. The on-site energy of a single boson at site $\ell$ is denoted by $\hbar (\omega_0 + \omega_\ell)$, where $\hbar \omega_0$ is the mean value calculated over the $L$ sites and $\hbar \omega_\ell$ measures the local deviation from the mean. We assume that $\omega_\ell$ are sampled from the uniform random distribution $\mathcal{U}(-D, D)$, where $D$ is the strength of the disorder. The third term in the Hamiltonian, $\hat{H}_U$, describes the attractive on-site interaction between the bosons, with $U > 0$ the strength of the interaction. Finally, the fourth term, $\hat{H}_J$, models the nearest-neighbour interaction between the sites, with $J > 0$ the hopping frequency.

Since the Hamiltonian commutes with the total number operator $\hat{N} = \sum_{\ell = 1}^L \hat{n}_\ell$, we can study separately the manifolds of fixed eigenvalues $N$ of $\hat{N}$. Since $\hat{H}_M \propto \hat{N}$, it is simply a constant when $N$ is fixed, and thus we can neglect it in our analysis.

\section{Open chain}
Let us first study the open chain, which was analysed in the main text.

\subsection{Localised phase}\label{subsec:open_localised}
Let us set the hopping frequency $J$ to zero and consider the Hamiltonian
\begin{equation}
\hat{H}_0 = \hat{H}_D + \hat{H}_U.
\end{equation}
The eigenstates of $\hat{H}_0$ at fixed $N$ are given by the tensor product states $\ket{n_1, \ldots, n_L} \equiv \ket{n_1} \otimes \cdots \otimes \ket{n_L}$ with $n_\ell$ excitations at site $\ell$ and $\sum_{\ell = 1}^L n_\ell = N$. The corresponding eigenenergies are 
\begin{equation*}
E_{n_1, \ldots, n_L} / \hbar = \sum_{\ell = 1}^{L} \omega_\ell n_\ell - \frac{U}{2} \sum_{\ell = 1}^{L} n_\ell (n_\ell - 1).
\end{equation*}
Using the facts that $\omega_\ell \geq \min_m \omega_m$ (by definition) and $\sum_{\ell = 1}^L n_\ell^2 \leq \left(\sum_{\ell = 1}^L n_\ell \right)^2$ (expanding the r.h.s.), we see that the ground state of the system is a state completely localized to the site $\ell_0$ with the lowest on-site energy $\omega_\ell$,
\begin{equation}
\ket{E^{(0)}} = \ket{n_{\ell_0} = N}, \label{eq.open_loc0}
\end{equation}
where $\ell_0 = \argmin{\ell}{\omega_\ell}$. The energy of the ground state is
\begin{equation}
E^{(0)} / \hbar = \omega_{\ell_0} N - \frac{U}{2}N(N-1).
\end{equation}
We assume here that the minimum is unique, that is, the ground state is non-degenerate.

\begin{figure*}  
    \centering
    \includegraphics[width=0.95\textwidth]{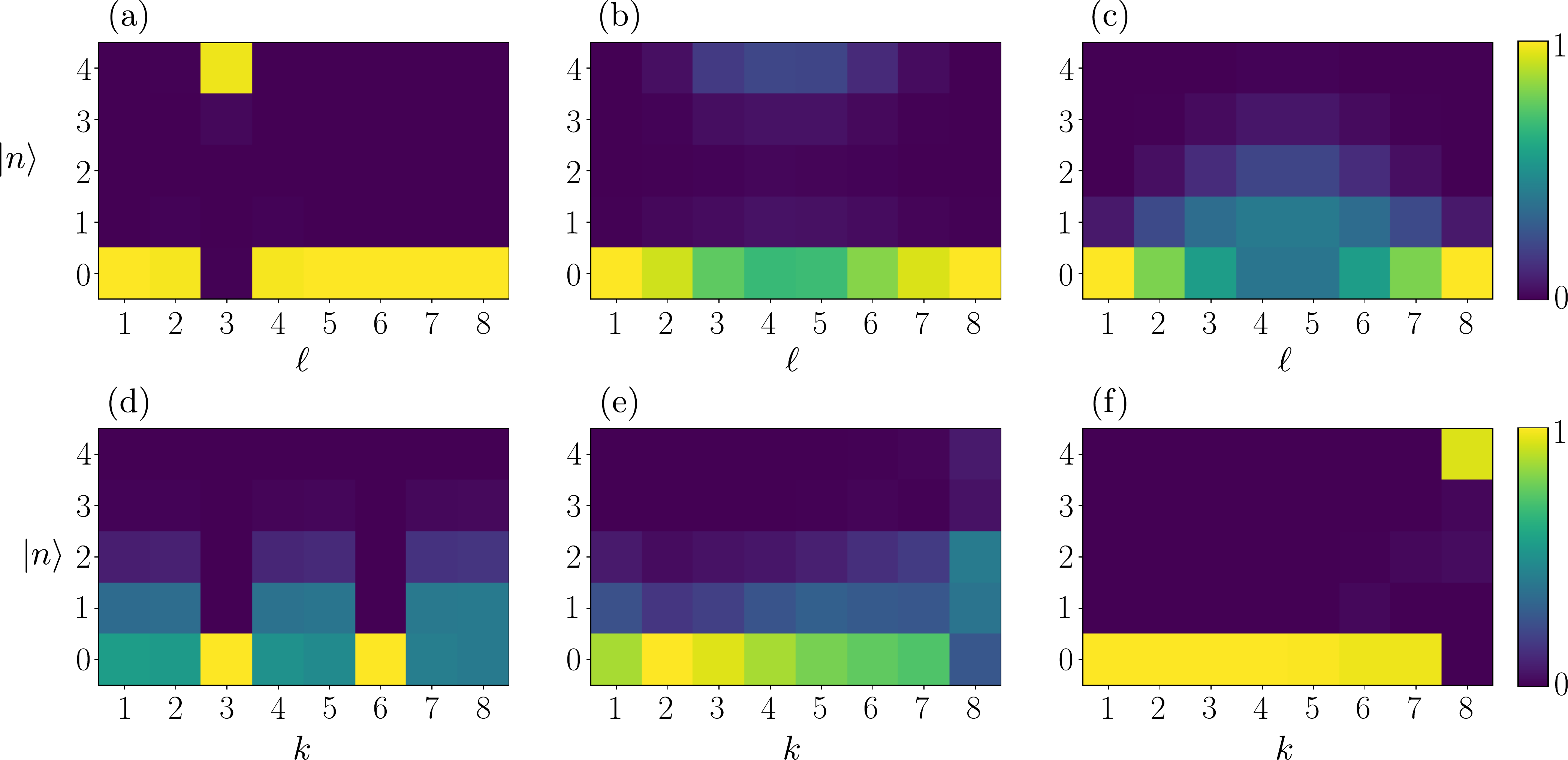}
    \caption{\label{fig.states_open}The occupation density of the open chain model in the position space $p_{n\ell}=\braket{n|\hat \rho_\ell|n}$, (a)--(c), as a function of the transmon site $\ell$ and the Fock state $\ket{n}$, and the occupation density in the reciprocal space $p_{nk}=\braket{n|\hat \rho_k|n}$, (d)--(f), as a function of the reciprocal mode $k$ and the corresponding Fock state $\ket{n}$ for (a),(d)~the~localized state, (b),(e)~the~W~state, and (c),(f)~the~superfluid~state at the scaled hopping frequency $\tau = 0.05$, $0.15$, and $1$, respectively, and at a single disorder realization at the scaled disorder strength $\delta =0.33\cdot10^{-3}$. The local density matrix $\hat \rho_\ell$ is calculated by tracing over all the other sites: $\hat \rho_\ell=\Tr_{\{i\neq \ell\}}\left(\ket{\psi}\bra{\psi}\right)$, where $\ket{\psi}$ is the full quantum ground state, and similarly for the reciprocal space. }

    \label{fig:1}
\end{figure*}

Let us then treat the hopping interaction $\hat{H}_J$ as a small perturbation to $\hat{H}_0$ using the standard Rayleigh-Schrödinger perturbation theory \cite{sakurai2017}, where we expand the ground state and the ground-state energy in powers of $J$ as $\ket{E} = \ket{E^{(0)}} + \ket{E^{(1)}} + \ldots$ and $E = E^{(0)} + E^{(1)} + \ldots$. The first-order correction to the ground-state energy is given by
\begin{equation}
E^{(1)} / \hbar = \bra{E^{(0)}} \hat{H}_J / \hbar \ket{E^{(0)}} = 0.
\end{equation}
The first-order correction to the ground state is given by
\begin{align}
&\ket{E^{(1)}}=\\ &\sum_{\substack{n_1, \ldots, n_L = 1 \\ n_1 + \cdots + n_L = N \\ n_{\ell_0} \neq N}}^{N} \ket{n_1, \ldots, n_L} \frac{\bra{n_1, \ldots, n_L} \hat{H}_J / \hbar \ket{n_{l_0} = N}}{(E^{(0)} - E_{n_1, \ldots, n_L}) / \hbar} . \notag 
\end{align}
Since the unperturbed ground state is completely localised, the only two states accessible via the perturbation $\hat{H}_J$ are $\ket{n_{\ell_0} = N - 1, n_{\ell_0 \pm 1} = 1}$. 
We thus obtain
\begin{align}
\ket{E^{(1)}} = J \sqrt{N} & \left(  \frac{\ket{n_{\ell_0 - 1} = 1, n_{\ell_0} = N - 1}}{\omega_{\ell_0} - \omega_{\ell_0 - 1} - U(N - 1)} \right.  \notag \\ & \quad \left. + \frac{\ket{n_{\ell_0} = N - 1, n_{\ell_0 + 1} = 1}}{\omega_{\ell_0} - \omega_{\ell_0 + 1} - U(N - 1)}  \right). \label{eq:open_loc_1st_ord_corr_state}
\end{align}
If it so happens that the state is localised at either one of the boundaries ($\ell_0 = 1$ or $\ell_0 = L$), we simply interpret $\ket{n_{0} = 1, n_1 = N - 1} = \ket{n_L = N - 1, n_{L + 1} = 1} = 0$.

Finally, the second-order correction to the ground-state energy is given by
\begin{equation}
E^{(2)} / \hbar = \bra{E^{(0)}} \hat{H}_J / \hbar \ket{E^{(1)}}.
\end{equation}
Using the result above, we obtain
\begin{align}
E^{(2)} / \hbar = J^2 N & \left( \frac{1}{\omega_{\ell_0} - \omega_{\ell_0 - 1} - U(N - 1)} \right. \notag \\ & \qquad \left.+ \frac{1}{\omega_{\ell_0} - \omega_{\ell_0 + 1} - U(N - 1)}  \right).
\end{align}
Again, this needs to be interpreted so that any term with $\omega_{1 - 1}$ or $\omega_{L+1}$ vanishes.

Since each different realisation of $\omega_\ell$:s produces a different result for the ground state energy $E \approx E^{(0)} + E^{(1)} + E^{(2)}$, we are more interested in its disorder average. In general, the disorder average of a function $f = f(\omega_1, \ldots, \omega_L)$ can be calculated as
\begin{equation}
\langle f \rangle_D = \int_{-D}^{D} d \omega_1 \cdots \int_{-D}^{D} d \omega_L \frac{f(\omega_1, \ldots, \omega_L)}{(2D)^L}.
\end{equation}
Since $E$ involves the minimum of $\omega_\ell$, we split the integration region $[-D, D]^L$ into $L$ distinct partitions $\omega_\ell \leq \omega_m$, $m \neq \ell$. A straightforward integration then yields
\begin{equation}
\varepsilon = - \frac{1}{2} - \delta \frac{L-1}{L+1} - 2 \tau^2 \frac{L-1}{L} \sum_{n = 1}^\infty \frac{(-1)^n L}{(n+1) (n+L)} (2 \delta)^n,
\end{equation}
where $\varepsilon = \langle E \rangle_D / \hbar U N (N - 1)$, $\tau = J / U(N - 1)$, and $\delta = D / U(N-1)$. The infinite series has a closed-form expression in terms of the natural logarithm, but we are mainly interested in the region where $2 \delta$ is small, and thus the series expansion is more convenient. The exact ground state of the localized phase is visualized in Fig.~\ref{fig.states_open}(a),(d) through the occupation density in the position and the reciprocal space, and the comparison with the analytic approximation of Eqs.~\eqref{eq.open_loc0} and~\eqref{eq:open_loc_1st_ord_corr_state} is shown in Fig.~\ref{fig.contours}(a). 

\subsection{W phase}\label{subsec:open_W}
Let us set both the hopping frequency $J$ and the disorder strength $D$ (and thus all $\omega_\ell$) to zero and consider the Hamiltonian
\begin{equation}
\hat{H}_0 = \hat{H}_U.
\end{equation}
We see from the results above that the ground state of $\hat{H}_0$ is $L$-fold degenerate and belongs to a manifold spanned by the localised states $\ket{n_{\ell} = N}$, $\ell = 1, \ldots, L$. The ground-state energy is given by
\begin{equation}
E^{(0)} / \hbar = - \frac{U}{2}N(N-1), \label{eq:open_W_E_0}
\end{equation}
but the exact form of the ground state is left undetermined.

If we now let $J > 0$, we can determine the shape of the ground state superposition in the limit $J \to 0$ using the standard degenerate perturbation theory \cite{sakurai2017}. However, unlike in the usual textbook examples, it is not enough to consider just first-order perturbations. Instead, one must compute the corrections up to the $L$th or the $N$th order, whichever is the lowest, as we shall see below.

Defining the projection operator $\hat{P}_0 = \sum_{\ell = 1}^L \ket{n_\ell = N} \bra{n_\ell = N}$ to the manifold spanned by the degenerate ground states of $\hat{H}_0$ (we will refer to this simply as the degenerate manifold), and its complement $\hat{P}_1 = \hat{I} - \hat{P}_0$ with $\hat{I}$ being the identity operator, the full eigenvalue problem of $\hat{H}_0 + \hat{H}_J$ can be written as \cite{sakurai2017}
\begin{align}
&\bigg\{ E - E^{(0)} - \hat{P}_0 \hat{H}_J \hat{P}_0  \label{eq:open_W_degenerate_perturbation_eq_0}  \\ \notag  & - \hat{P}_0 \hat{H}_J \left[\hat{P}_1 \left(E - \hat{H}_0 - \hat{H}_J  \right) \hat{P}_1 \right]^{-1} \hat{H}_J \hat{P}_0 \bigg\} \hat{P}_0 \ket{E} = 0, \\
&\hat{P}_1 \ket{E} = \left[\hat{P}_1 \left(E - \hat{H}_0 - \hat{H}_J  \right) \hat{P}_1 \right]^{-1} \hat{H}_J \hat{P}_0 \ket{E}. \label{eq:open_W_degenerate_perturbation_eq_1}
\end{align}
Here, $\ket{E}$ is the exact ground state and $E$ is the exact ground-state energy, which we again expand in powers of $J$ as $\ket{E} = \ket{E^{(0)}} + \ket{E^{(1)}} + \ldots$ and $E = E^{(0)} + E^{(1)}~+~\ldots$. Equation (\ref{eq:open_W_degenerate_perturbation_eq_0}) is an eigenvalue problem within the degenerate manifold, and determines the shape of the solution therein, in particular the correct form of the ground state $\ket{E^{(0)}}$ of $\hat{H}_0$. After solving Eq.\ (\ref{eq:open_W_degenerate_perturbation_eq_0}), Eq.\ (\ref{eq:open_W_degenerate_perturbation_eq_1}) can be used to calculate the solution outside the degenerate manifold.

\begin{figure*}
    \centering
    \includegraphics[width=0.95\linewidth]{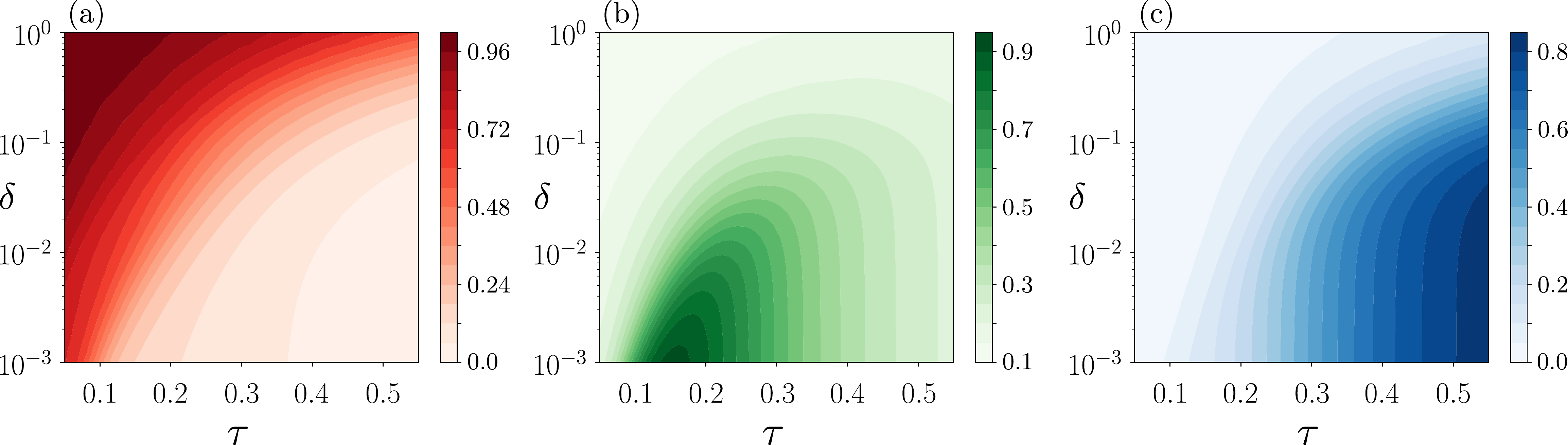}
    \caption{\label{fig.contours}The overlap $|\braket{\psi|\psi'}|^2$ of the exact ground state $\ket{\psi}$ and the analytically solved $\ket{\psi'}$ of the open chain model as a function of the scaled hopping frequency $\tau$ and disorder strength $\delta$ for (a) the localized phase $\ket{\psi'}=\ket{\psi_{\rm loc}}$ of Eqs.~\eqref{eq.open_loc0} and~\eqref{eq:open_loc_1st_ord_corr_state}, (b) the W phase $\ket{\psi_{\rm W}}$ of Eq.~\eqref{eq.w0N}, and (c) the superfluid phase $\ket{\psi_{\rm SF}}$ of Eqs.~\eqref{eq.sf0},~\eqref{eq.sf1D}~and~\eqref{eq.sf1U}. The overlap was  computed for an open chain of length $L= 8$ with the total number of bosons $N= 4$, and averaged over \num{1000} disorder realisations }
\end{figure*}

Noting that $\hat{P}_0 \hat{H}_J \hat{P}_0$ vanishes identically, and expanding the inverse of $\hat{P}_1 (E - \hat{H}_0 - \hat{H}_J  ) \hat{P}_1$ using Neumann series, Eq.\ (\ref{eq:open_W_degenerate_perturbation_eq_0}) can be written as (still no approximations)
\begin{equation}
\left(E - E^{(0)} - \sum_{i = 2}^\infty \hat{K}_i \right) \hat{P}_0 \ket{E} = 0, \label{eq:open_W_degenerate_perturbation_eq_0_mod}
\end{equation}
where
\begin{align}
\hat{K}_i = \hat{P}_0 \hat{H}_J & \left\{\left[ \hat{P}_1 \left(E - \hat{H}_0 \right) \hat{P}_1 \right]^{-1} \hat{P}_1 \hat{H}_J \hat{P}_1 \right\}^{i - 2} \notag \\ & \qquad \quad \times  \left[ \hat{P}_1 \left(E - \hat{H}_0 \right) \hat{P}_1 \right]^{-1} \hat{H}_J \hat{P}_0. \label{eq:open_W_K_l}
\end{align}
The operators $\hat{K}_i$ behave essentially like $(\hat{H}_J)^i$. They are explicitly of order $i$ in $J$, although higher-order terms emerge when expanding $E$ in powers of $J$. Working in the basis $\ket{n_1 = N}, \ldots, \ket{n_L = N}$, we see that $\hat{K}_i$ are diagonal if $i < N$, since it takes at least $N$ moves to transfer $N$ excitations from one site to another. For odd $i$, all the diagonal terms vanish, since it is impossible to move excitations localised to one particular site by an odd number of times and still return to the original state. If $i $ is even, $i  = 2 m$, the diagonal of $\hat{K}_i$ is of the form $\mathrm{diag}\{k_1, \ldots, k_m, k_{m+1}, k_{m+1}, \ldots, k_{m+1}, k_{m+1}, k_m, \ldots, k_1 \}$, where $k_j > k_{j+1}$. This is explained by the fact that for sites farther away from the boundaries, there are more ways to move excitations around before hitting the boundary. Since each such route contributes to $\hat{K}_i$ with a negative weight, the result follows. Finally, $\hat{K}_N$ is tridiagonal, with equal super- and subdiagonal elements. This follows from the fact that there is exactly one way to transfer $N$ excitations from one site to another site with $N$ moves. In particular, the target site has to be one of the nearest neighbours.

With these observations in mind, we can proceed with solving Eq.\ (\ref{eq:open_W_degenerate_perturbation_eq_0_mod}) for $\ket{E^{(0)}}$. We work our way up order by order, up to order $N$. All odd-order corrections to the ground state energy vanish, since $\hat{K}_i = 0$ for odd $i$. In particular,
\begin{equation}
E^{(1)} / \hbar = 0.
\end{equation}
Each even-order term $\hat{K}_{2m}$ up to order $N-1$ reduces the dimension of the ground state manifold by dropping out the $m$ sites closest to each of the boundaries, since these are higher in energy. Thus, if $N$ is high enough compared to $L$, only one or two sites participate in the ground state. To be more precise, this happens if $L \leq 2 \lceil N/2 \rceil$, where $\lceil x \rceil$ denotes the floor of $x$, i.e., the smallest integer greater than or equal to $x$. For odd $L$, this completely determines the unperturbed ground state,
\begin{equation}
\ket{E^{(0)}} = \ket{n_{(L+ 1)/2} = N}.
\end{equation}
For even $L$, the two middlemost sites have equal energy up to $N-1$st order, but at $N$th order the degeneracy is lifted, and we find that the unperturbed ground state is
\begin{equation}
\ket{E^{(0)}} = \frac{1}{\sqrt{2}} \left(\ket{n_{L/2} = N} + (-1)^N \ket{n_{L/2 + 1} = N} \right)
\end{equation}
by solving a simple $2 \times 2$ eigenvalue problem.

If, on the other hand, $L > 2 \lceil N/2 \rceil$, then the unperturbed ground state is a linear combination of the $L_d = L + 2 - 2 \lceil N/2 \rceil$ middlemost states $\ket{n_{\ell + \ell_s} = N}$, where $l_s = \lceil N/2 \rceil - 1$ and $\ell = 1, \ldots, L_d$. The exact form of this linear combination is determined at $N$th order. If $N$ is odd, the diagonal of $\hat{K}_N$ vanishes, and the problem reduces to finding the eigenvalues and -vectors of a symmetric tridiagonal Toeplitz matrix with non-negative coefficients. This is a well-known problem, with the solution \cite{losonczi1992eigenvalues}

\begin{equation}
\ket{E^{(0)}} = \sqrt{\frac{2}{L_d + 1}} \sum_{\ell = 1}^{L_d} (-1)^\ell \sin \left(\frac{\pi \ell}{L_d + 1} \right) \ket{n_{\ell + \ell_s} = N}.
\end{equation}
If $N$ is even, the diagonal of $\hat{K}_N$ does not vanish, and we need to solve the eigenproblem of a matrix of the form
\begin{equation}
\begin{pmatrix}
a + c & b & 0 & 0 & 0 & \hdots & 0 \\
b & a & b & 0 & 0 & \cdots & 0 \\
0 & b & a & b & 0 & \cdots & 0 \\
0 & 0 & b & a & b & \cdots & 0 \\
\vdots & & & \ddots & \ddots & \ddots &  \\
0 & \cdots & 0 & 0 & b & a & b \\
0 & \cdots & 0 & 0 & 0 & b & a + c
\end{pmatrix},
\end{equation}
where $a < 0$,
\begin{align}
b &= - U N (N-1) \frac{(N-1)^{N-1}}{(N-1)!} \tau^N, \\
c &= U N (N-1) \tau^N,
\end{align}
and $\tau = J / U(N-1)$. Note that the value of $a$ has no effect on the eigenstates, it simply shifts the eigenvalues. Also note that $0 \leq c \leq |b|$. A fully closed-form solution is available if $c = 0$ or $c = |b|$ \cite{losonczi1992eigenvalues}. If $N=2$, then $c = |b|$ and the ground state is given by
\begin{equation}
\ket{E^{(0)}} = \sqrt{\frac{2}{L_d}} \sum_{\ell = 1}^{L_d} \sin \left(\frac{\pi (2\ell - 1)}{2 L_d} \right) \ket{n_{\ell + \ell_s} = N}. 
\end{equation}
If $N > 2$, the ratio $c / |b|$ is always quite small compared to unity, attaining the value $2/9$ for $N = 4$ and decreasing exponentially with increasing $N$. Thus, a good approximation for the ground state is obtained by simply setting $c = 0$. In this case, we have
\begin{equation}
\ket{E^{(0)}} = \sqrt{\frac{2}{L_d + 1}} \sum_{\ell = 1}^{L_d} \sin \left(\frac{\pi \ell}{L_d + 1} \right) \ket{n_{\ell + \ell_s} = N}. \label{eq.w0N}
\end{equation}
For completeness, if $0 < c < |b|$, the ground state is of the form
\begin{align}
\ket{E^{(0)}} = \sqrt{\frac{2}{A L_d + B}} \sum_{\ell = 1}^{L_d} & \left\{c \sin [(\ell - 1) \theta ] - b \sin(\ell \theta) \right\}\notag \\ & \hspace{1.25cm}\times\ket{n_{\ell + \ell_s} = N},
\end{align}
where $\theta$ is the smallest positive solution to the equation
\begin{equation}
\cos \left(\frac{L_d + 1}{2} \theta \right) + \left|\frac{c}{b} \right| \cos \left(\frac{L_d - 1}{2} \theta \right) = 0
\end{equation}
and $A = b^2 + c^2 - 2 b c \cos \theta$, $B = b^2 - c^2$. 

Going to $N+1$st order, we find that for odd $N$, the first order correction $\hat{P}_0 \ket{E^{(1)}}$ is a linear combination of the same states as $\ket{E^{(0)}}$, but orthogonal to this (choosing the usual normalisation $\braket{E^{(0)} | E} = 1$). In the case of even $N$, which is the case we are more interested in, $\hat{P}_0 \ket{E^{(1)}}$ vanishes. This is due to the fact that $\hat{K}_{N+1}$ vanishes, since $N+1$ is odd.

In the space orthogonal to the degenerate manifold, the first-order correction to the ground state can be calculated using Eq.\ (\ref{eq:open_W_degenerate_perturbation_eq_1}). Each state $\ket{n_\ell = N}$ in the linear combination attains a correction identical to the one we calculated for the localised phase, see Eq.\ (\ref{eq:open_loc_1st_ord_corr_state}).

Finally the second-order correction to the ground-state energy is given by
\begin{equation}
E^{(2)} / \hbar =  \bra{E^{(0)}} \hat{H}_J / \hbar \ket{E^{(1)}} = - \frac{2 J^2 N}{U(N-1)}.
\end{equation}
The expected value of the ground state energy $E \approx E^{(0)} + E^{(1)} + E^{(2)}$ is thus given by
\begin{equation}
\varepsilon = -\frac{1}{2} - 2 \tau^2,
\end{equation}
where $\varepsilon = \langle E \rangle_D / \hbar U N (N - 1)$ and $\tau = J / U(N - 1)$. The exact ground state of the W phase is visualized in Fig.~\ref{fig.states_open}(b),(e) through the occupation density in the position and the reciprocal space, and the comparison with the analytic approximation of Eq.~\eqref{eq.w0N} is shown in Fig.~\ref{fig.contours}(b). 

\subsection{Effect of disorder on the W phase}\label{subsec:open_W_disorder}
Above, we discussed how the hopping term $\hat{H}_J$ lifts the $L$-fold degeneracy of the ground state of the Hamiltonian $\hat{H}_U$, fixing the correct form of the superposition of the states $\ket{n_\ell = N}$ to the W state. In the analysis, we neglected the disorder altogether. Here, we briefly discuss the consequences of disorder on the W phase.

Treating both $\hat{H}_D$ and $\hat{H}_J$ as small perturbations to $\hat{H}_U$, the unperturbed ground state $\ket{E^{(0)}}$ of $\hat{H}_U$ can be solved from Eq.\ (\ref{eq:open_W_degenerate_perturbation_eq_0})  after making the replacement $\hat{H}_J \to \hat{H}_D + \hat{H}_J$. Note that unlike $\hat{P}_0 \hat{H}_J \hat{P}_0$, the term $\hat{P}_0 \hat{H}_D \hat{P}_0 = N \sum_{\ell = 1}^L \omega_\ell \ket{n_\ell = N} \bra{n_\ell = N}$ does not vanish. This immediately shows us that if $D \sim J$, the disorder is the dominant term, present already at first order in the perturbation theory, making the localised phase energetically favourable compared to the W phase. 

Let us again denote $\tau = J / U(N - 1)$ and $\delta = D / U(N - 1)$, and assume that $L > 2 \lceil N/2 \rceil$. If $\delta \sim \tau^n$ for $n \geq 2$, the disorder first appears at order $n$ in the perturbation theory, using $\tau$ (or $J$) as the expansion parameter. On the other hand, we saw above that the exact form of the W state is determined at order $N$. This leads us to the following conclusion. If $n > N$, the disorder only slightly modifies the W phase. If $n < N$, the disorder is the dominant perturbation, and thus the localised phase is favoured (although the hopping term may still reduce the possible sites of localisation to ones sufficiently far away from the boundaries, depending on the value of $n$). If $n = N$, both $\hat{H}_D$ and $\hat{H}_J$ contribute more or less equally, and we are at the transition region between the two phases. Therefore, roughly speaking, the W phase is stable when $\delta \lesssim \tau^N$. Otherwise, the localised phase is favoured.

Let us consider the transition region in a bit more detail. Assuming $\delta \sim \tau^N$, the ground state of the system is some superposition of the states $\ket{n_{\ell + \ell_s} = N}$, $\ell = 1, \ldots, L_d$, where $\ell_s = \lceil N/2 \rceil - 1$ and $L_d = L - 2 \ell_s$ (see subsection \ref{subsec:open_W}). The form of this linear combination is determined by the matrix $\hat{P}_0 \hat{H}_D \hat{P}_0 + \hat{K}_N^{(0)}$, the ground state of which we need to solve within this subspace. Here, we have denoted $\hat{K}_N^{(0)} = \hat{K}_N(E = E^{(0)})$. Now, the ground state of $\hat{P}_0 \hat{H}_D \hat{P}_0$ is a state localised to the site $\ell_0$ with the lowest value of $\omega_\ell$. On the other hand, $\hat{K}_N^{(0)}$ couples neighbouring sites with strength $\hbar U N (N-1) (N-1)^{N-1} \tau^N / (N-1)!$. When this coupling increases enough compared to the energy difference between neighbouring sites, the average value of which is $2 \hbar D N / 3$, the localised state starts to spread considerably. Increasing the coupling even further, the ground state starts to approach the W state. Comparing the coupling strength with the energy difference, we find that the phase boundary obeys the formula
\begin{equation}
\delta = (C \tau)^N,
\end{equation}
where
\begin{equation}
C = \sqrt[N]{\frac{3}{2} A \frac{(N-1)^{N-1}}{(N-1)!}}
\end{equation}
and $A$ is a constant close to unity. Note that, by Stirling's approximation for factorials, $C \to e$ as $N \to \infty$.

\subsection{Superfluid phase}\label{subsec:open_superfluid}
Let us set the on-site interaction strength $U$ and the disorder strength $D$ to zero, and consider the Hamiltonian
\begin{equation}
\hat{H}_0 = \hat{H}_J.
\end{equation}
One way to find the eigenstates and eigenenergies of $\hat{H}_0$ is to first consider the Heisenberg equations of motion for $\hat{a}_\ell$,
\begin{equation*}
\frac{d \hat{a}_\ell}{dt} = i \left[\hat{H}_0 / \hbar, \hat{a}_\ell \right] = -i J \hat{a}_{\ell - 1} - i J \hat{a}_{\ell + 1}
\end{equation*}
This is a simple tridiagonal Toeplitz system, which can be diagonalised with a transformation \cite{losonczi1992eigenvalues}
\begin{equation}
\hat{a}_\ell = \sqrt{\frac{2}{L+1}} \sum_{k = 1}^L \sin \left(\frac{\pi \ell k}{L+1} \right) \hat{c}_k.
\end{equation}
In terms of the reciprocal space creation and annihilation operators $\hat{c}_k^\dagger$ and $\hat{c}_k$, the Hamiltonian can be written as
\begin{equation}
\hat{H}_0 / \hbar = 2 J \sum_{k = 1}^L \cos \left(\frac{\pi k}{L+1} \right) \hat{c}_k^\dagger \hat{c}^{}_k.
\end{equation}
The total number operator is still of the same form as before, $\hat{N} = \sum_{k = 1}^L \hat{c}_k^\dagger \hat{c}^{}_k=\sum_{k=1}^L\hat \eta_k$. 

The eigenstates of $\hat{H}_0$ at fixed $N$ are given by the tensor product states $\ket{\eta_1, \ldots, \eta_L} \equiv \ket{\eta_1} \otimes \cdots \otimes \ket{\eta_L}$ with $\eta_k$ excitations at the $k$th reciprocal mode and $\sum_{k = 1}^L \eta_k = N$. The corresponding eigenenergies are 
\begin{equation}
E_{\eta_1, \ldots, \eta_L} / \hbar = 2 J \sum_{k = 1}^L \cos \left(\frac{\pi k}{L+1} \right) \eta_k.
\end{equation}
Since cosine is a decreasing function in the interval $[0, \pi]$, we see that the ground state of the system is a state completely localized to mode $L$,
\begin{equation}
\ket{E^{(0)}} = \ket{\eta_L = N}, \label{eq.sf0}
\end{equation}
The energy of the ground state is
\begin{equation}
E^{(0)} / \hbar = - 2 J N \cos \left(\frac{\pi}{L+1} \right).
\end{equation}

Let us then treat the disorder $\hat{H}_D$ and the on-site interaction $\hat{H}_U$ as small perturbations to $\hat{H}_0$. Expressed in the reciprocal space, these are given by
\begin{align}
\hat{H}_D/ \hbar &=  \sum_{\ell, j, k = 1}^{L} \frac{2 \omega_\ell}{L + 1} \sin \left(\frac{\pi \ell j}{L + 1} \right) \sin \left(\frac{\pi \ell k}{L + 1} \right) \hat{c}_j^\dagger \hat{c}^{}_k,\\
\hat{H}_U / \hbar &= - \frac{U}{2} \left(\frac{2}{L + 1} \right)^2 \sum_{\ell, j, k, m, n = 1}^{L} \sin \left(\frac{\pi \ell j}{L + 1} \right) \\ 
\times & \sin \left(\frac{\pi \ell k}{L + 1} \right) \sin \left(\frac{\pi \ell m}{L + 1} \right) \sin \left(\frac{\pi \ell n}{L + 1} \right) \hat{c}_j^\dagger \hat{c}_k^\dagger \hat{c}_m \hat{c}_n. \notag 
\end{align}
The first-order correction to the ground-state energy is given by
\begin{equation}
E^{(1)} / \hbar = \bra{E^{(0)}} \hat{H}_D / \hbar + \hat{H}_U / \hbar \ket{E^{(0)}}.
\end{equation}
A straightforward calculation yields
\begin{equation}
E^{(1)} / \hbar = \frac{2 N}{L + 1} \sum_{\ell = 1}^L \omega_\ell \sin^2 \left(\frac{\pi \ell}{L+1} \right) - \frac{3 U N (N-1)}{4 (L+1)}.
\end{equation}
The first-order correction to the ground state is given by
\begin{align}
&\ket{E^{(1)}} = \label{eq.1stsf} \\ & \notag \sum_{\substack{\eta_1, \ldots, \eta_L = 1 \\ \eta_1 + \cdots + \eta_L = N \\ \eta_L \neq N}}^{N} \ket{\eta_1, \ldots, \eta_L} \frac{\bra{\eta_1, \ldots, \eta_L} \hat{H}_D  + \hat{H}_U \ket{\eta_L = N}}{(E^{(0)} - E_{\eta_1, \ldots, \eta_L})}.
\end{align}
The only states giving a non-vanishing contribution are $\ket{\eta_k = 1, \eta_L = N - 1}$ for $k = 1, \ldots, L - 1$, $\ket{\eta_k = 2, \eta_L = N - 2}$ for $k = 1, \ldots, L - 1$, and $\ket{\eta_k = 1, \eta_{k + 2} = 1, \eta_L = N - 2}$ for $k = 1, \ldots, L - 3$. A straightforward, albeit a little tedious, calculation shows that $\ket{E^{(1)}} = \ket{E^{(1)}_D} + \ket{E^{(1)}_U}$, where

\begin{widetext}
\begin{align}
\ket{E^{(1)}_D} & = -\frac{\sqrt{N}}{J (L + 1)} \sum_{k = 1}^{L-1}  \frac{\ket{\eta_k = 1, \eta_L = N-1}}{\cos \left(\frac{\pi}{L + 1} \right) + \cos \left(\frac{\pi k}{L + 1} \right)} \sum_{\ell = 1}^{L} \omega_\ell \sin \left(\frac{\pi \ell k}{L + 1} \right) \sin \left(\frac{\pi \ell L}{L + 1} \right) \label{eq.sf1D}\\
\ket{E^{(1)}_U} &= \frac{U \sqrt{N (N-1)}}{8 J (L+1)} \Bigg\{\sum_{k = 1}^{L - 1} \frac{\sqrt{2} (2 + \delta_{k, 1}) \ket{\eta_k = 2, \eta_L = N - 2}}{2 \left[\cos \left(\frac{\pi}{L + 1} \right) + \cos \left(\frac{\pi k}{L + 1} \right) \right]}  \notag \\ 
&\hspace*{1cm}- \frac{2 \sqrt{N - 1} \ket{\eta_{L - 2} = 1, \eta_L = N - 1}}{\cos \left(\frac{\pi}{L + 1} \right) + \cos \left(\frac{\pi (L - 2)}{L + 1} \right)}- \sum_{k = 1}^{L - 3} \frac{2 \ket{\eta_k = 1, \eta_{k+2} = 1, \eta_L = N - 2}}{2 \cos \left(\frac{\pi}{L + 1} \right) + \cos \left(\frac{\pi k}{L + 1} \right) + \cos \left(\frac{\pi (k+2)}{L + 1} \right)} \Bigg\}. \label{eq.sf1U}
\end{align}

Finally, the second-order correction to the ground-state energy is given by
\begin{equation}
E^{(2)} / \hbar = \bra{E^{(0)}} \hat{H}_D / \hbar + \hat{H}_U / \hbar \ket{E^{(1)}}.
\end{equation}
Using the result above, we obtain
\begin{equation}
\begin{split}
E^{(2)} / \hbar = &- \frac{2 N}{J (L+1)^2} \sum_{k = 1}^{L-1} \sum_{\ell_1 = 1}^L \sum_{\ell_2 = 1}^L \omega_{\ell_1} \omega_{\ell_2} \frac{\sin \left(\frac{\pi \ell_1 k}{L + 1} \right) \sin \left(\frac{\pi \ell_2 k}{L + 1} \right) \sin \left(\frac{\pi \ell_1 L}{L + 1} \right) \sin \left(\frac{\pi \ell_2 L}{L + 1} \right)}{\cos \left(\frac{\pi}{L + 1} \right) + \cos \left(\frac{\pi k}{L + 1} \right)} \\
&- \frac{U^2 N (N - 1)}{32 J (L+1)^2} \left[\frac{15 + 4 L (L + 2)}{6 \cos \left(\frac{\pi}{L + 1} \right)} + \frac{6 \cos^2 \left(\frac{\pi}{L + 1} \right) - 5}{\sin^2 \left(\frac{\pi}{L + 1} \right) \cos \left(\frac{\pi}{L + 1} \right) } + \frac{4 (N - 1)}{\cos \left(\frac{\pi}{L + 1} \right) - \cos \left(\frac{3 \pi}{L + 1} \right) } \right] \\
&- \frac{N(N - 1)}{J (L+1)^2} \frac{\sum_{\ell = 1}^L \omega_\ell \sin \left(\frac{\pi \ell L}{L + 1} \right) \sin \left(\frac{\pi \ell (L - 2)}{L + 1} \right)}{\cos \left(\frac{\pi}{L + 1} \right) + \cos \left(\frac{\pi (L - 2)}{L + 1} \right)}.
\end{split}
\end{equation}
\begin{figure*}
    \centering
    \includegraphics[width=0.95\textwidth]{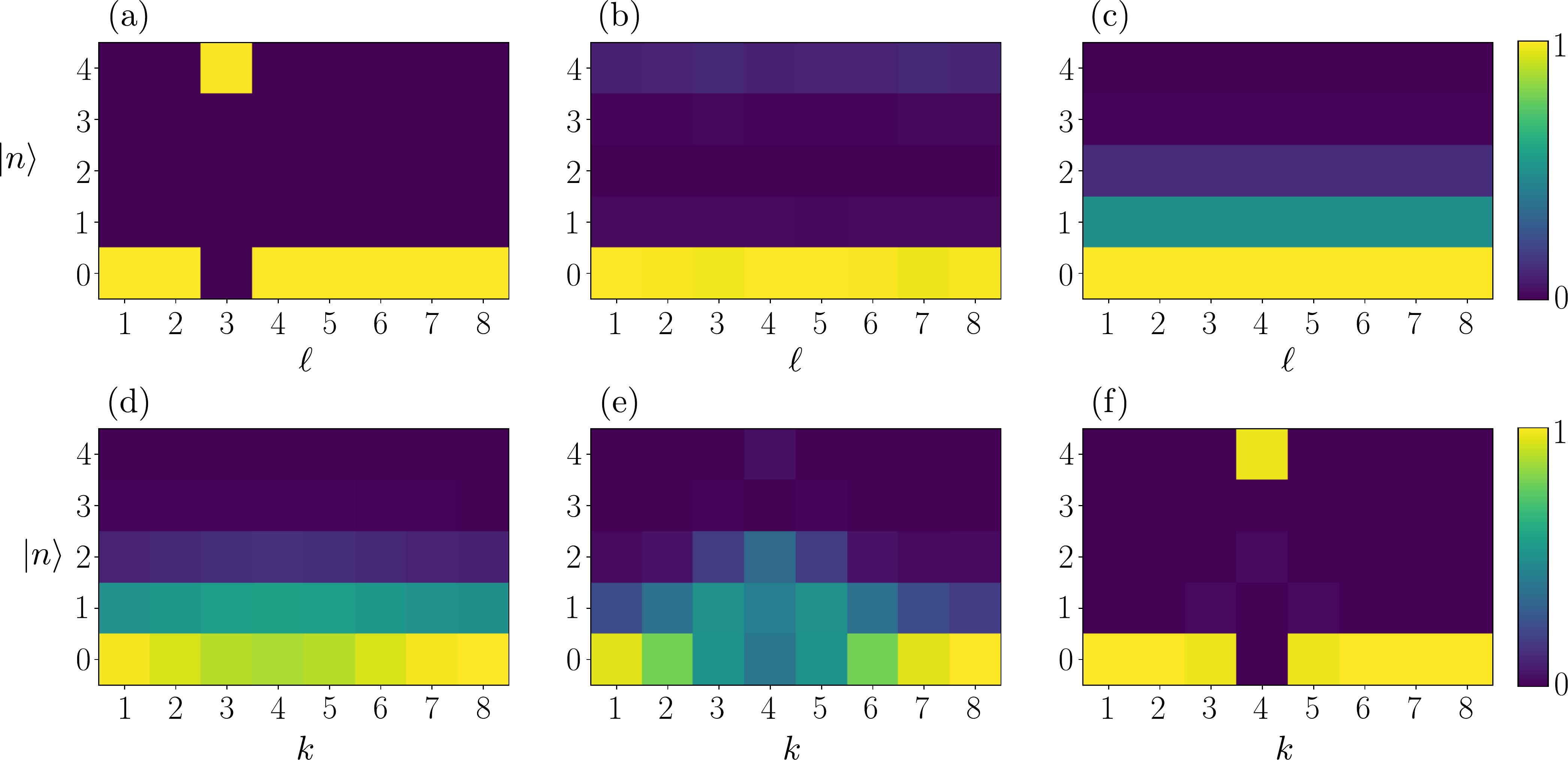}
    \caption{\label{fig.states_periodic}The occupation density of the periodic chain model in the position space $p_{n\ell}=\braket{n|\hat \rho_\ell|n}$, (a)--(c), as a function of the transmon site $\ell$ and the Fock state $\ket{n}$, and the occupation density in the reciprocal space $p_{nk}=\braket{n|\hat \rho_k|n}$, (d)--(f), as a function of the reciprocal mode $k$ and the corresponding Fock state $\ket{n}$ for (a),(d)~the~localized state, (b),(e)~the~W~state, and (c),(f)~the~superfluid~state at the scaled hopping frequency $\tau = 0.05$, $0.15$, and $1$, respectively, and at a single disorder realization at the scaled disorder strength $\delta =0.33\cdot10^{-3}$. The local density matrix $\hat \rho_\ell$ is calculated by tracing over all the other sites: $\hat \rho_\ell=\Tr_{\{i\neq \ell\}}\left(\ket{\psi}\bra{\psi}\right)$, where $\ket{\psi}$ is the full quantum ground state, and similarly for the reciprocal space.}
\end{figure*}We are again interested in the expected value of the ground state energy $E \approx E^{(0)} + E^{(1)} + E^{(2)}$. A straightforward calculation shows that
\begin{equation}
\begin{split}
\varepsilon = &- 2 \tau \cos \left(\frac{\pi}{L + 1} \right) - \frac{3}{4 (L+1)} - \frac{\delta^2}{24 \tau (L+1)} \frac{5 \cos^2 \left(\frac{\pi}{L+1} \right) + 1}{\cos \left(\frac{\pi}{L+1} \right) \sin^2 \left(\frac{\pi}{L+1} \right)} \\
&- \frac{1}{32 \tau (L+1)^2 (N-1)} \Bigg[\frac{15 + 4 L (L + 2)}{6 \cos \left(\frac{\pi}{L+1} \right)} + \frac{6 \cos^2 \left(\frac{\pi}{L+1} \right) - 5}{\sin^2 \left(\frac{\pi}{L+1} \right) \cos \left(\frac{\pi}{L+1} \right) } + \frac{4 (N - 1)}{\cos \left(\frac{\pi}{L+1} \right) - \cos \left(\frac{3 \pi}{L+1} \right) } \Bigg],
\end{split}
\end{equation}
\end{widetext}
where $\varepsilon = \langle E \rangle_D / \hbar U N (N - 1)$, $\tau = J / U(N - 1)$, and $\delta = D / U(N-1)$. The exact ground state of the superfluid phase is visualized in Fig.~\ref{fig.states_open}(c),(f) through the occupation density in the position and the reciprocal space, and the comparison with the analytic approximation of Eqs.~\eqref{eq.sf0},~\eqref{eq.sf1D}~and~\eqref{eq.sf1U} is shown in Fig.~\ref{fig.contours}(c).

\section{Periodic chain}
For a periodic chain, the analysis is very much similar to the case of an open chain, albeit simpler in general.

\subsection{Localised phase}
Let us set the hopping frequency $J$ to zero and consider the Hamiltonian
\begin{equation}
\hat{H}_0 = \hat{H}_D + \hat{H}_U.
\end{equation}
The results for the ground state and ground-state energy are exactly the same as the ones obtained for an open chain in subsection \ref{subsec:open_localised}, with the slight simplification that, since there are no boundaries in the chain, the sites $\ell = 1$ and $\ell = L$ are now equivalent to other sites.

\begin{figure*}
    \centering
    \includegraphics[width=0.95\linewidth]{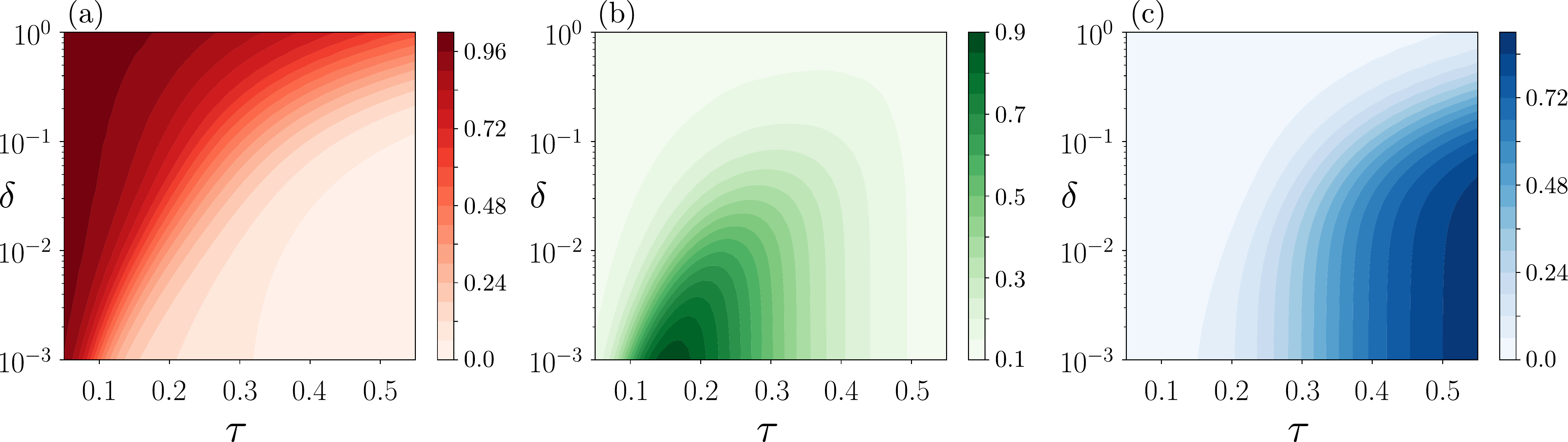}
    \caption{\label{fig.contours_periodic} The overlap $|\braket{\psi|\psi'}|^2$ of the exact ground state $\ket{\psi}$ and the analytically solved  ground states $\ket{\psi'}$ of the periodic model as a function of the scaled hopping frequency $\tau$ and disorder strength $\delta$ for (a) the localized phase $\ket{\psi'}=\ket{\psi_{\rm loc}}$ of Eqs.~\eqref{eq.open_loc0} and~\eqref{eq:open_loc_1st_ord_corr_state}, (b) the W phase $\ket{\psi_{\rm W}}$ of Eq.~\eqref{eq.w0N_periodic}, and (c) the superfluid phase $\ket{\psi_{\rm SF}}$ of Eqs.~\eqref{eq.sf0_periodic},~\eqref{eq.sf1D_periodic}~and~\eqref{eq.sf1U_periodic}. The overlap was computed for an periodic chain of length $L= 8$ with the total number of bosons $N= 4$, and averaged over \num{1000} disorder realisations }
\end{figure*}

\subsection{W phase}
Let us set both the hopping frequency $J$ and the disorder strength $D$ to zero and consider the Hamiltonian
\begin{equation}
\hat{H}_0 = \hat{H}_U.
\end{equation}
Like in the case of an open chain, the ground state of $\hat{H}_0$ is $L$-fold degenerate, with the states $\ket{n_\ell = N}$ all having the same energy of Eq.(\ref{eq:open_W_E_0}). 

The degeneracy is again lifted if $J > 0$. Following the analysis of subsection \ref{subsec:open_W}, we need to solve Eq.\ (\ref{eq:open_W_degenerate_perturbation_eq_0_mod}), where $\hat{K}_i$ are defined in Eq.\ (\ref{eq:open_W_K_l}). Due to the absence of boundaries, the following analysis is now much simpler.

Again, the operators $\hat{K}_i$ are of order $i$ in $J$. They are diagonal for $i < N$ and the diagonal elements vanish for odd $i$. Unlike in an open chain, however, the diagonal elements for even $i$ are all equal. This means that nothing is resolved until order $N$ in perturbation theory, and no special cases have to be considered depending on the relative size of $N$ and $L$. In $N$th order, the ground state is determined by $\hat{K}_N$. Neglecting the constant diagonal term which does not affect the shape of ground state, $\hat{K}_N$ is of the form
\begin{equation}
b
\begin{pmatrix}
0 & 1 & 0 & 0 & 0 & \hdots & 1 \\
1 & 0 & 1 & 0 & 0 & \cdots & 0 \\
0 & 1 & 0 & 1 & 0 & \cdots & 0 \\
0 & 0 & 1 & 0 & 1 & \cdots & 0 \\
\vdots & & & \ddots & \ddots & \ddots & \\
0 & \cdots & 0 & 0 & 1 & 0 & 1 \\
1 & \cdots & 0 & 0 & 0 & 1 & 0
\end{pmatrix},
\end{equation}
where
\begin{equation}
b = U N (N-1) (-1)^{N-1} \frac{(N-1)^{N-1}}{(N-1)!} \tau^N
\end{equation}
and $\tau = J / U(N-1)$. This is diagonal in the Fourier space. For even $N$, the coefficient $b$ is less than zero, leading to the ground state
\begin{equation}
\ket{E^{(0)}} = \sqrt{\frac{1}{L}} \sum_{\ell = 1}^{L} \ket{n_\ell = N}. \label{eq.w0N_periodic}
\end{equation}
For odd $N$, $b > 0$ and the ground state depends on the parity of $L$. If $L$ is even, the ground state is
\begin{equation}
\ket{E^{(0)}} = \sqrt{\frac{1}{L}} \sum_{\ell = 1}^{L} (-1)^\ell \ket{n_\ell = N}. 
\end{equation}
If $L$ is odd, the ground state is doubly degenerate, with the states
\begin{equation}
\ket{E^{(0)}_\pm} = \sqrt{\frac{1}{L}} \sum_{\ell = 1}^{L} (-1)^\ell e^{\pm i \pi \ell / L} \ket{n_\ell = N} 
\end{equation}
having equal energies.

The rest of the results are identical to the ones calculated in subsection \ref{subsec:open_W}. 

\subsection{Effect of disorder on the W phase}
Disorder affects the W phase in a periodic chain in the same way as in an open chain, see subsection \ref{subsec:open_W_disorder}.

\subsection{Superfluid phase}
Let us set the on-site interaction strength $U$ and the disorder strength $D$ to zero, and consider the Hamiltonian
\begin{equation}
\hat{H}_0 = \hat{H}_J.
\end{equation}
Since the chain is $L$-periodic and $\hat{H}_J$ couples nearest neighbours with equal strength, it is natural to switch to reciprocal space representation using the Fourier transform
\begin{equation}
\hat{a}_\ell = \frac{1}{\sqrt{L}} \sum_{k = 1}^L \exp \left(\frac{2 \pi i \ell k}{L} \right) \hat{b}_k.
\end{equation}
In terms of the reciprocal space creation and annihilation operators $\hat{b}_k^\dagger$ and $\hat{b}_k$, the Hamiltonian can be written as
\begin{equation}
\hat{H}_0 / \hbar = 2 J \sum_{k = 1}^L \cos \left(\frac{2 \pi k}{L} \right) \hat{b}_k^\dagger \hat{b}^{}_k.
\end{equation}
The total number operator is still of the same form as before, $\hat{N} = \sum_{k = 1}^L \hat{b}_k^\dagger \hat{b}_k$. 

The eigenstates of $\hat{H}_0$ at fixed $N$ are given by the tensor product states $\ket{\eta_1, \ldots, \eta_L} \equiv \ket{\eta_1} \otimes \cdots \otimes \ket{\eta_L}$ with $\eta_k$ excitations at the $k$th reciprocal mode and $\sum_{k = 1}^L \eta_k = N$. The corresponding eigenenergies are 
\begin{equation}
E_{\eta_1, \ldots, \eta_L} / \hbar = 2 J \sum_{k = 1}^L \cos \left(\frac{2 \pi k}{L} \right) \eta_k.
\end{equation}
For simplicity, let us consider only the case of even $L$. Since cosine obtains its minimum at $\pi$, we see that the ground state of the system is a state completely localized to mode $L/2$,
\begin{equation}
\ket{E^{(0)}} = \ket{\eta_{L/2} = N}, \label{eq.sf0_periodic}
\end{equation}
and the energy of the ground state is
\begin{equation}
E^{(0)} / \hbar = - 2 J N.
\end{equation}
Note that if we were to consider the case of odd $L$, the ground state would be degenerate, with the states $\ket{\eta_{(L-1)/2} = N}$ and $\ket{\eta_{(L+1)/2} = N}$ having equal energies.

Let us then treat the disorder $\hat{H}_D$ and the on-site interaction $\hat{H}_U$ as small perturbations to $\hat{H}_0$. Expressed in the reciprocal space, these are given by
\begin{align}
\hat{H}_D / \hbar &= \frac{1}{L} \sum_{\ell, j, k = 1}^{L} \omega_\ell e^{2 \pi i \ell (k-j)/L} \hat{b}_j^\dagger \hat{b}^{}_k, \\
\hat{H}_U / \hbar &= - \frac{U}{2 L^2} \sum_{\ell, j, k, m, n = 1}^{L} e^{2 \pi i \ell (m+n-j-k)/L}\hat{b}_j^\dagger \hat{b}_k^\dagger \hat{b}^{}_m \hat{b}^{}_n.
\end{align}
The first-order correction to the ground-state energy is given by
\begin{equation}
E^{(1)} / \hbar = \bra{E^{(0)}} \hat{H}_D / \hbar + \hat{H}_U / \hbar \ket{E^{(0)}}.
\end{equation}
A straightforward calculation yields
\begin{equation}
E^{(1)} / \hbar = \frac{N}{L} \sum_{\ell = 1}^L \omega_\ell - \frac{U N (N-1)}{2 L}.
\end{equation}

The first-order correction to the ground state $\ket{\eta_{L/2}=N}$ is given by Eq.~\eqref{eq.1stsf}. 

The only states giving a non-vanishing contribution are $\ket{\eta_k = 1, \eta_{L/2} = N - 1}$ for $k \neq L/2$, $\ket{\eta_{L/2} = N-2, \eta_L = 2}$, and $\ket{\eta_k = 1, \eta_{L/2} = N-2, \eta_{L-k} = 1}$ for $k = 1, \ldots, L/2 - 1$. A straightforward calculation shows that $\ket{E^{(1)}} = \ket{E^{(1)}_D} + \ket{E^{(1)}_U}$, where
\begin{align}
\ket{E^{(1)}_D} &= -\frac{\sqrt{N}}{2 J L} \sum_{\substack{k = 1 \\ k \neq L/2}}^{L}  \frac{\ket{\eta_k = 1, \eta_{L/2} = N-1}}{1 + \cos \left(\frac{2 \pi k}{L} \right)} \notag \\ &  \hspace*{1.5cm} \times \sum_{\ell = 1}^{L} \omega_\ell (-1)^\ell \exp \left(-\frac{2 \pi i \ell k}{L} \right) \label{eq.sf1D_periodic}\\
\ket{E^{(1)}_U} & = \frac{U \sqrt{N (N-1)}}{4 J L} \Bigg\{\frac{\ket{\eta_{L/2} = N-2, \eta_L = 2}}{2 \sqrt{2}} \label{eq.sf1U_periodic} \\
& +\sum_{k = 1}^{L/2 - 1} \frac{\ket{\eta_k = 1, \eta_{L/2} = N-2, \eta_{L-k} = 1}}{1 + \cos \left(\frac{2 \pi k}{L} \right)} \Bigg\}. \notag 
\end{align}

Finally, the second-order correction to the ground-state energy is given by
\begin{equation}
E^{(2)} / \hbar = \bra{E^{(0)}} \hat{H}_D / \hbar + \hat{H}_U / \hbar \ket{E^{(1)}}.
\end{equation}
Using the result above, we obtain
\begin{align}
&E^{(2)} / \hbar = - \frac{U^2 N (N-1) (L^2-1)}{48 J L^2} \\
&- \frac{N}{2 J L^2} \sum_{\substack{k = 1 \\ k \neq L/2}}^{L} \sum_{\ell_1 = 1}^L \sum_{\ell_2 = 1}^L \omega_{\ell_1} \omega_{\ell_2} (-1)^{\ell_1 + \ell_2} \frac{e^{ 2 \pi i k (\ell_1 - \ell_2)/L}}{1 + \cos \left(\frac{2 \pi k}{L} \right)}. \notag 
\end{align}
Calculating the expected value of the ground state energy $E \approx E^{(0)} + E^{(1)} + E^{(2)}$ over different realisations of $\omega_\ell$, we obtain
\begin{equation}
\begin{split}
\varepsilon = &- 2 \tau - \frac{1}{2 L} - \frac{\delta^2 (L^2-4)}{36 \tau L} - \frac{L^2-1}{48 \tau (N-1) L^2},
\end{split}
\end{equation}
where $\varepsilon = \langle E \rangle_D / \hbar U N (N - 1)$, $\tau = J / U(N - 1)$, and $\delta = D / U(N-1)$. 

Figure~\ref{fig.states_periodic} shows the visualizations of the localized phase, the W phase and the superfluid phase of the periodic model through the occupation density in the position and the reciprocal space. Figure~\ref{fig.contours_periodic} shows the overlap between the analytic approximations and the numerically computed exact ground state for the periodic chain model.

\section{Inverse participation ratios}
\begin{figure*}[h!]
    \centering
    \includegraphics[width=0.95\linewidth]{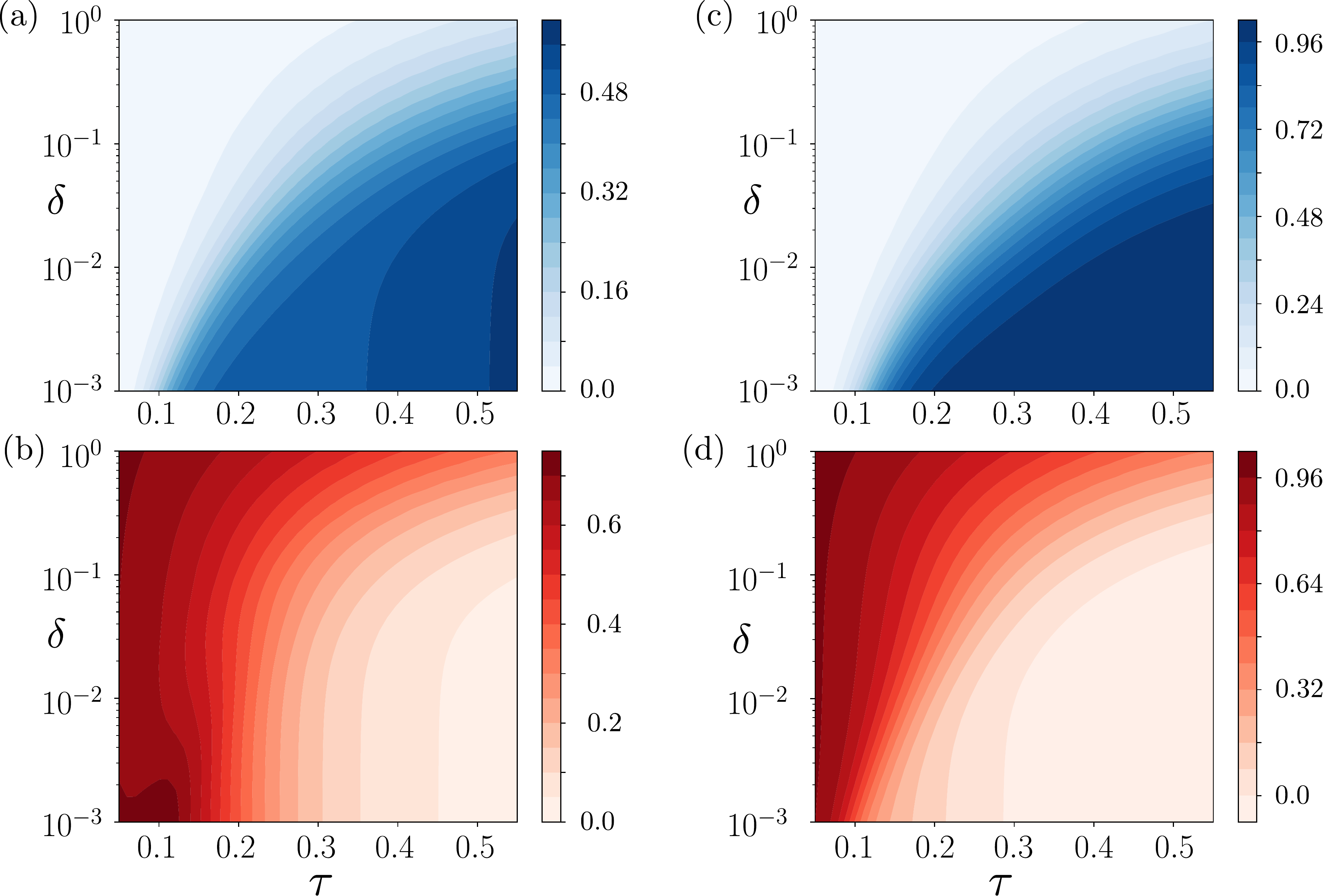}
    \caption{\label{fig.ipr_open_periodic}The spatial $\mathcal P_{\rm s}$ (a),(c) and reciprocal $\mathcal P_{\rm r}$ (b),(d) inverse participation ratios of the quantum ground state of the disordered attractive Bose--Hubbard model as a function of the scaled hopping frequency $\tau$ and disorder strength $\delta$. The ground states were numerically computed for an open chain (a)--(b) and a periodic chain (c)--(d) of length $L = 8$ with the total number of bosons $N = 4$, and averaged over 1000 disorder realisations.}
\end{figure*}

The spatial $\mathcal P_{\rm s}$ and reciprocal $\mathcal P_{\rm r}$ inverse participation ratios are defined as 
\begin{align}
  \mathcal{P}_{\rm s} & = \frac{1}{L - 1}\left(\frac{N^2}{\sum_{m = 1}^L|\braket{\psi|\hat n^{\rm s}_m|\psi}|^2} - 1\right),\label{eq.Ps} \\
  \mathcal{P}_{\rm r} & = \frac{1}{L - 1}\left(\frac{N^2}{\sum_{m = 1}^L|\braket{\psi|\hat n^{\rm r}_m|\psi}|^2} - 1\right).\label{eq.Pr}
\end{align}
The Fig.~2(a) of the main text shows both inverse participation ratios as a function of the scaled hopping frequency $\tau=J/U(N-1)$ at two values of the scaled disorder strength $\delta = D/U(N-1)$ for an open chain geometry. In Fig.~\ref{fig.ipr_open_periodic} (a)--(b), we show these open chain inverse participation ratios in the whole parameter range. The Fig.~2(c) of the main text is made by combining these two as a single figure. Figures~\ref{fig.ipr_open_periodic} (c)--(d) show the corresponding inverse participation ratios in a periodic chain model.

\bibliography{refs.bib}{}